\documentclass[aps,prd,10pt,nofootinbib,noeprint,amsmath,amssymb,superscriptaddress,preprintnumbers,showkeys]{revtex4-2}

\usepackage[utf8]{inputenc} % Accents.
\usepackage[T1]{fontenc} % Fonts.
\usepackage[english]{babel} % English typography.
\usepackage{xcolor} % Colors management.
\usepackage{amsmath,amsfonts,amssymb,amsthm,braket,dsfont,comment,leftindex,bbm} % Mathematical symboles.
\usepackage{mathrsfs}
\usepackage{graphicx} % Picture management.
\usepackage{ragged2e} % pour mettre des sous-figures
\usepackage{tikz} % Diagrams production.
\usepackage{multirow,afterpage,tabularray} % Table management.
\usepackage{fancyhdr,orcidlink} % Footnotes & headers.
\usepackage[capitalise]{cleveref} % Improved handling of numeric citations
\usepackage{enumitem} % Enumeration customization.
\usepackage{titlesec} % Title edition.
\usepackage{lipsum}
\usepackage{listings} % Insert code with syntax.
\usepackage{cancel}
\usepackage{hyperref}

\hypersetup{
    colorlinks=true,
    urlcolor=blue,
    linkcolor=blue,
    citecolor=blue}

\renewcommand{\d}{\mathrm{d}}

\titleformat{\subsubsection}
   {\normalfont\fontsize{9pt}{11pt}\selectfont\bfseries}% apparence commune au titre et au numéro
   {\thesubsubsection.}% apparence du numéro
   {1em}% espacement numéro/texte
   {\centering}% apparence du titre
\titleformat*{\subsubsection}{\normalsize\itshape}

\begin{document}

\title{Tidal deformations of general-relativistic multifluid compact stars}

\author{\surname{Ethan} Carlier \orcidlink{0009-0008-6076-1422}}
\email[E-mail: ]{ethan.carlier@ulb.be}
\affiliation{Institute of Astronomy and Astrophysics, Université Libre de Bruxelles, CP 226, Boulevard du Triomphe, B-1050 Brussels, Belgium}
\affiliation{Brussels Laboratory of the Universe, Belgium}
\affiliation{Leuven Gravity Institute, KU Leuven, Celestijnenlaan 200D box 2415, 3001 Leuven, Belgium}

\author{\surname{Nicolas} Chamel \orcidlink{0000-0003-3222-0115}}
\email[E-mail: ]{nicolas.chamel@ulb.be}
\affiliation{Institute of Astronomy and Astrophysics, Université Libre de Bruxelles, CP 226, Boulevard du Triomphe, B-1050 Brussels, Belgium}
\affiliation{Brussels Laboratory of the Universe, Belgium}
\affiliation{Leuven Gravity Institute, KU Leuven, Celestijnenlaan 200D box 2415, 3001 Leuven, Belgium}

\date{\today}

\begin{abstract}

Over the past decade, gravitational-wave astronomy has opened a new window onto the extreme states of matter inside compact stars. At some point during the inspiral of a binary system, each star starts to experience adiabatic tides, characterized by tidal deformabilities. The dominant gravitoelectric quadrupolar tidal deformability, first measured with the famous GW170817 binary neutron-star merger event, has already constrained the dense-matter equation of state. With the advent of third-generation detectors, tidal deformabilities are expected to be inferred with much higher precision, potentially revealing subleading tidal contributions. This motivates the development of more accurate compact-star models that incorporate richer microphysics. With this in mind, we move beyond the commonly adopted perfect‑fluid approximation and model compact stars through a multifluid framework that captures a broader set of microscopic properties. In this work, we present the fully general-relativistic description of adiabatic tidal deformations of compact stars composed of an arbitrary number of interacting fluids, using Carter's multifluid variational formalism. A distinctive feature of this approach is the presence of nondissipative mutual entrainment between fluid species. We derive the hydrostatic equilibrium equations for multifluid configurations, along with the perturbed equations governing stationary gravitoelectric and gravitomagnetic tidal responses of arbitrary order. We then investigate how entrainment modifies the corresponding tidal deformabilities. Using an analytical representation of the multifluid equation of state, we show that entrainment leaves adiabatic tidal responses unchanged and therefore produces no measurable effect on the gravitational-wave signal emitted during the inspiral long before the excitation of internal mode resonances. We subsequently discuss two specific applications: superfluid neutron stars, and dark matter admixed compact stars.

\end{abstract}


\maketitle

\section{Introduction and motivation}
\label{intro}

Compact stars, including white dwarfs, neutron stars, and more exotic hypothetical objects, are unique astrophysical laboratories. They contain matter subject to extreme physical conditions~\cite{glendenning_compact_1996, shapiro_black_1983}. The internal constitution of white dwarfs is relatively well understood, and consists of a dense Coulomb plasma of atomic nuclei (primarily carbon and oxygen) immersed in a relativistic electron gas. Neutron stars, on the other hand, exhibit a much richer structure: as their matter is crushed to even higher average baryon number densities deep enough in their crust, some neutrons become unbound and with increasing density nuclei progressively dissolve into a dense liquid of nucleons and leptons. Although neutron stars have been extensively observed since their fortuitous discovery in 1967, the properties of their dense matter remains very uncertain. As a result, a large diversity of theoretical models for the equation of state have been developed, leading to significant variations in the predicted macroscopic properties of neutron stars, such as their masses, radii, and tidal deformabilities (see~\cite{oertel_equations_2017, burgio_neutron_2021} for comprehensive reviews).

Beyond this canonical picture of compact stars, exotic scenarios have been proposed. At densities exceeding a few times that found in atomic nuclei, additional particle species may appear altering the composition of neutron-star cores. In particular, numerous studies have explored the possible presence of hyperons~\cite{chatterjee_hyperons_2016,vidana_hyperons_2021,sedrakian_heavy_2023} and even deconfined quarks~\cite{annala_strongly_2023}. It has been speculated that so-called ``strange stars'' might be entirely made of strange quark matter~\cite{glendenning_compact_1996,weber_strange_2005,haensel_neutron_2007}. White dwarfs could also host a quark core giving rise to a distinct class of so-called ``strange dwarfs''~\cite{glendenning_strange_1995, glendenning_possible_1995}. A variety of other exotic configurations has been discussed in the literature. In the context of this work, we highlight dark matter admixed compact stars~\cite{bramante_dark_2024}. In addition to the diversity of particle degrees of freedom, compact stars may host a wide variety of quantum phases. In neutron stars, neutrons are expected to form a superfluid both in the inner crust and in the core, while protons are believed to become superconducting in the core~\cite{dean_pairing_2003,sedrakian_superfluidity_2019}. If hyperons are present, these strange baryons may also undergo superfluid transitions~\cite{sedrakian_superfluidity_2019, tu_1s_0_2022}. Furthermore, if deconfined quark matter exists inside compact stars, it is likely to appear in a color-superconducting phase, and a rich variety of such phases has been extensively investigated~\cite{buballa_njl-model_2005,alford_color_2008,anglani_crystalline_2014}.

In recent years, the emergence of multimessenger astronomy has opened new ways of investigating the internal composition of compact stars~\cite{abbott_multi-messenger_2017}. In particular, observations of gravitational waves from binary neutron-star mergers, combined with electromagnetic counterparts, have already begun to place constraints on the dense-matter equation of state~\cite{margalit_multi-messenger_2019, dietrich_multimessenger_2020}. Future third-generation gravitational-wave detectors, such as the Einstein Telescope and Cosmic Explorer, will drastically improve sensitivity to tidal effects and will make the merger and post-merger dynamics accessible, thereby offering unprecedented access to the microphysics at supranuclear densities~\cite{abac_science_2025, evans_horizon_2021}. We focus here on the inspiral stage of binary coalescence, during which each star is only slightly deformed by the external tidal field of its companion. As long as the orbital frequency remains well below the characteristic frequencies of internal modes, no stellar oscillations are excited and the tidal responses are fully characterized by the (adiabatic) tidal Love numbers~\cite{hinderer_tidal_2008, flanagan_constraining_2008, dietrich_interpreting_2021}. These quantities enter directly into waveform approximants and can therefore be potentially inferred from gravitational-wave data analyses~\cite{ligo_scientific_collaboration_and_virgo_collaboration_gw170817_2017, most_new_2018}. Approximants within the post-Newtonian framework incorporate tidal corrections up to 7.5PN order, associated with the quadrupolar and octupolar electric Love numbers (typically denoted $k_2$ and $k_3$, respectively) as well as the quadrupolar magnetic Love number (denoted by $j_2$)~\cite{henry_tidal_2020}. Effective-one-body models such as \texttt{TEOBResumS}~\cite{nagar_time-domain_2018} and \texttt{SEOBNRv4T}~\cite{hinderer_effects_2016, steinhoff_dynamical_2016} include tidal interactions incorporating higher-order Love numbers up to $k_4$. Phenomenological approximants like \texttt{NRTidalv*} also account for tidal effects~\cite{dietrich_closed-form_2017, dietrich_improving_2019, colleoni_new_2025}. Since Love numbers depend sensitively on the underlying microphysics, their extraction from gravitational-wave observations provide constraints on dense-matter properties. The Einstein Telescope collaboration anticipates that next-generation detectors will measure tidal deformabilities with an order-of-magnitude gain in precision relative to the design sensitivity of the LIGO-Virgo-KAGRA network, potentially enabling the detection of subleading tidal contributions~\cite{abac_science_2025, jimenez_forteza_impact_2018}. This prospect motivates the development of increasingly accurate theoretical models of tidal interactions, together with more realistic descriptions of matter. Indeed, systematic uncertainties arising from theoretical modelling are expected to constitute the dominant source of error in data analyses for the next generation of ground-based gravitational-wave observatories \cite{abac_science_2025}.

Most relativistic calculations of adiabatic Love numbers of compact stars assume a single, barotropic, cold perfect fluid~\cite{hinderer_tidal_2008,damour_relativistic_2009,binnington_relativistic_2009}. Various generalizations have been explored. 
Elasticity has been incorporated to account for the solid crust of neutron stars \cite{penner_tidal_2011, gittins_tidal_2020,pereira_tidal_2020}, the crystallized core of white dwarfs \cite{perot_tidal_2022,tang_oscillations_2023}, the solid hadronic layers of strange dwarfs~\cite{perot_unmasking_2023}, and a possible crystalline color superconducting quark core in various compact stars~\cite{lau_tidal_2017,perot_role_2023,dong_lovec_2024,dong_new_2025}; while the effect is completely negligible for neutron stars, it can substantially modify the tidal responses of white dwarfs and compact stars containing a solid quark core. Thermal effects induced by tidal friction during the late inspiral of binary neutron stars have been investigated in \cite{kanakis-pegios_thermal_2022} and found to be very small. Superfluidity in cold neutron stars has been considered in \cite{char_relativistic_2018, datta_effect_2020, passamonti_dynamical_2022, yeung_i-love-q_2021, aranguren_revisiting_2023} within a two-fluid model. However, these studies lead to contradictory conclusions. On the one hand, the numerical results obtained in Refs.~\cite{char_relativistic_2018, datta_effect_2020} indicate that superfluidity can change tidal deformabilities by as much as 20\% due to non-dissipative mutual entrainment effect. On the other hand, the authors of Ref.~\cite{passamonti_dynamical_2022} argued that if the unperturbed configuration is in beta equilibrium, then the dominant tidal deformability is independent of entrainment effect and reduces to the barotropic one-fluid case. The effects of dark matter inside compact stars on tidal deformations have been investigated for a wide range of dark-matter models~\cite{mukhopadhyay_quark_2016, mukhopadhyay_compact_2017, leung_tidal_2022, das_dark_2022, barbat_comprehensive_2024}. A two-fluid model is also adopted; however, possible entrainment between baryonic matter and dark matter has, to the best of our knowledge, not been considered in these studies.

In this paper, we revisit the role of entrainment in multifluid compact stars using Carter’s relativistic variational formalism~\cite{carter_covariant_1989}. In particular, we discuss the physically relevant cases of superfluid neutron stars and dark matter admixed compact stars. Carter’s formalism is remarkably general and can describe various models of compact stars. The one-fluid limit reproduces the standard cold, barotropic, perfect-fluid model. The two-fluid formalism can describe: (1) a minimal description of cold superfluid neutron stars, consisting of a neutron superfluid and a second fluid representing all remaining species~\cite{langlois_differential_1998}; (2) finite-temperature compact stars, in which entropy is treated as an independent component~\cite{lopez-monsalvo_thermal_2010}; or (3) baryonic compact objects admixed with dark matter~\cite{ciarcelluti_have_2011, leung_tidal_2022}. A three-fluid model can, for example, account for (1) the presence of superfluid hyperons, or (2) the inclusion of temperature effects in superfluid neutron stars or dark matter admixed objects. This list is not exhaustive: quark matter or mixed phases may require an even larger number of fluids, and a detailed description of neutron-star superfluidity may involve up to seven components~\cite{rau_relativistic_2020}. Here, we consider a compact star composed of an arbitrary number $N$ of fluid species, subject to an external adiabatic tidal field. We demonstrate that one can obtain perturbation equations valid for any value of $N$, and we provide explicit expressions for $N = 1$ to $N = 3$. We then clarify the impact of the entrainment effect on the tidal deformabilities. We prove, in a very general setting, that entrainment does not affect adiabatic tidal responses. This implies that superfluidity, as well as entrainment between baryonic and dark matter, leave no imprint on the gravitational signal during the early inspiral. Throughout this work, we assume that all fluid species are electrically charge neutral. In principle, elasticity and dissipation can also be accommodated~\cite{andersson_relativistic_2021}, but these aspects will not be considered here.

The structure of the paper is as follows. In Sec.~\ref{multifluid}, we present an overview of Carter’s multifluid formalism in its most general formulation. In Sec.~\ref{equilibrium}, we apply this framework to a static and spherically symmetric (i.e., non-spinning) equilibrium configuration. Section~\ref{perturbation} discusses the stationary tidal perturbations in both the gravitoelectric and gravitomagnetic sectors. In Sec.~\ref{Love numbers}, we relate the fluid perturbations to the external tidal field and introduce the tidal Love numbers and tidal deformabilities. The impact of entrainment on the tidal responses is examined in Sec.~\ref{entrainement}. Finally, we conclude by summarizing our results and outlining possible directions for future work.

Throughout this manuscript, we adopt the following notations and conventions. The components of an arbitrary tensor, $p$ times covariant and $q$ times contravariant, are written as $A^{\mu_1 \dots \mu_p}{}_{\nu_1 \dots \nu_q}$. Greek indices $\alpha, \beta, \dots, \mu, \nu, \dots$ take the values $0$, $1$, $2$, and $3$. Latin indices $a, b, \dots, i, j, \dots$ are restricted to the spatial values $1$, $2$, and $3$. The Einstein summation convention is applied to spacetime indices: any repeated index is implicitly summed over. In addition to spacetime indices, tensorial quantities related to the fluids also carry labels indexing the fluid species. For this purpose, we use only the letters $\mathrm{x}, \mathrm{y}, \mathrm{z}$, and $\mathrm{w}$ in this typographic form. These labels are not subject to covariance (i.e., their position does not matter) and do not follow Einstein’s summation convention. The symmetric and antisymmetric parts of a tensor are defined, respectively, by $A_{(\mu\nu)} = (A_{\mu\nu}+A_{\nu\mu})/2$ and $A_{[\mu\nu]} = (A_{\mu\nu}-A_{\nu\mu})/2$. We use geometric units in which $c = G = 1$. The metric of flat Minkowski spacetime is $\eta_{\mu\nu} = \mathrm{diag}(-1, +1, +1, +1)$. For curvature-related quantities, we follow the Misner-Thorne-Wheeler conventions~\cite{misner_gravitation_1973}.

\section{Relativistic multifluid dynamics}
\label{multifluid}

We begin by briefly reviewing the general formalism used to describe a self-gravitating compact star composed of $N$ interacting, non-dissipative fluids in the  general-relativistic context. Our approach relies on the powerful and elegant variational principle developed extensively by Carter and his collaborators \cite{carter_covariant_1989, carter_convective_1991, carter_equation_1995, comer_hamiltonian_1993, carter_relativistic_1998, carter_relativistic_2001}. These works have been comprehensively reviewed in \cite{andersson_relativistic_2021}.

\subsection{Unconstrained action principle}

The fundamental quantities characterizing the macroscopic state of the star are the spacetime metric $g_{\mu\nu}$, and the density currents $n^\mu_\mathrm{x}$, where $\mathrm{x}~\in~\{1, 2, \dots, N\}$ labels the fluid species. The central ingredient of the relativistic multifluid formalism is the so-called master function $\Lambda(n^\mu_\mathrm{x}, g_{\mu\nu})$, which plays the role of the Lagrangian density in the action principle. To construct a scalar and isotropic Lagrangian, $\Lambda$ must depend only on scalar combinations of the metric and the currents. These scalars are defined as
\begin{equation}
    n^2_\mathrm{xy} = - n^\mu_\mathrm{x} n^\nu_\mathrm{y} g_{\mu\nu},
\end{equation}
so that $\Lambda = \Lambda(n^2_\mathrm{xy})$. When all fluids are comoving, $-\Lambda$ reduces to the total static energy density. In this sense, the master function can be viewed as a multifluid equation of state, encapsulating the microphysical properties of the system. The key difference from a static equation of state arises from the presence of coupled current terms in $\Lambda$.

The total action $\mathcal{S} = \mathcal{S}\left(n^\mu_\mathrm{x}, g_{\mu\nu}\right)$, consisting of the Einstein-Hilbert term and the matter contribution, reads
\begin{equation}
    \mathcal{S} = \mathcal{S}_\mathrm{EH} + \mathcal{S}_\mathrm{matter} = \frac{1}{16\pi}\int \sqrt{-g}\left(\mathcal{R}+16\pi \Lambda \right) \d^4x,
\end{equation}
where $\mathcal{R}$ is the Ricci scalar and $g$ the determinant of the metric. To obtain the equations of motion, the action is required to be stationary, i.e. $\delta\mathcal{S} = 0$. Because $\Lambda$ can be viewed either as a functional of $n^2_\mathrm{xy}$ or as a function of~$(n^\mu_\mathrm{x}, g_{\mu\nu})$, we may express its variation in two equivalent forms:
\begin{subequations}
    \begin{align}
        \delta \Lambda &= - \frac{1}{2}\sum_\mathrm{x,y}\mathcal{A}_\mathrm{xy} \Upsilon_\mathrm{xy}\delta n^2_\mathrm{xy} \label{delta Lambda 1} \\
        &= \sum_\mathrm{x} \pi_\alpha^\mathrm{x} \delta n^\alpha_\mathrm{x} + \frac{1}{2}\sum_\mathrm{x} \pi^\alpha_\mathrm{x} n^\mu_\mathrm{x} \delta g_{\alpha\mu}. \label{delta Lambda 2}
    \end{align}
    \label{delta Lambda}
\end{subequations}
The matrix elements $\mathcal{A}_\mathrm{xy}$ appearing in \eqref{delta Lambda 1} are defined by
\begin{equation}
    \mathcal{A}_\mathrm{xy} = - \frac{\partial \Lambda}{\partial n^2_\mathrm{xy}}.
    \label{A}
\end{equation}
We also introduced the shifted Kronecker symbol
\begin{equation}
    \Upsilon_{\mathrm{xy}} = \delta_{\mathrm{xy}} + 1 =
    \begin{cases}
        1 & \text{if } \mathrm{x} \neq \mathrm{y} \\
        2 & \text{if } \mathrm{x} = \mathrm{y}
    \end{cases} \ ,
\end{equation}
which ensures the correct counting of distinct and identical pairs. The covariant quantities
\begin{equation}
    \pi^\mathrm{x}_\alpha = \frac{\partial \Lambda}{\partial n^\alpha_\mathrm{x}} = \sum_\mathrm{y} \mathcal{A}_\mathrm{xy} \Upsilon_\mathrm{xy} n^\mathrm{y}_\alpha,
\end{equation}
are the momenta canonically and thermodynamically conjugate to the currents $n^\alpha_{\mathrm{x}}$. We notice from the expression of the momentum of a given fluid species~$\mathrm{x}$ that it is not aligned with the corresponding current. Instead, the momentum is in general a linear combination of the currents of all fluid species. In other words, each fluid component transfers part of its motion to the others. This phenomenon is known as the mutual entrainment effect, and it is encoded in the off-diagonal terms of the matrix elements~$\mathcal{A}_{\mathrm{xy}}$. The entrainment effect disappears for a given species~$\mathrm{x}$ when $\mathcal{A}_{\mathrm{xy}} = 0$ for all $\mathrm{y} \neq \mathrm{x}$. This occurs when the master function depends only on the diagonal current terms, i.e.\ $\Lambda = \Lambda(n^{2}_{\mathrm{xx}})$.

If one were to vary the action naively with respect to the fields $n^\alpha_{\mathrm{x}}$, equation~\eqref{delta Lambda 2} would imply $\pi^\mathrm{x}_{\alpha} = 0$, corresponding to fluids that carry neither energy nor momentum—an unphysical result. To avoid this issue, we impose the conservation of each flux,
\begin{equation}
    \nabla_\mu n^\mu_{\mathrm{x}} = 0,
    \label{conserved flux}
\end{equation}
and adopt the so-called convective variational principle. In this approach, one considers displacements of the fluid-particle worldlines of the form $x^\mu_\mathrm{x} \rightarrow x^\mu_\mathrm{x} + \xi^\mu_\mathrm{x}$. As a result, the action principle becomes constrained: the variations $\delta n^\alpha_{\mathrm{x}}$ are no longer independent of the metric variations $\delta g_{\mu\nu}$. To understand how these variations are related, one introduces the notion of Lagrangian variation.

\subsection{Eulerian and Lagrangian variations}

Up to now, when performing the variation of a given quantity $A$, we have meant the change of the field at the same spacetime point. This is what is called an Eulerian variation, denoted by $\delta A$, and it corresponds to a more macroscopic point of view. There exists, however, another way to consider variations: the Lagrangian variation~\cite{friedman_generic_1978}. The latter is more microscopic, in the sense that 
it measures the change of a quantity along the motion of a fluid element. More precisely, the Lagrangian displacements $\xi^\alpha_{\mathrm{x}}$ map a fluid element in the unperturbed configuration to the same element in the perturbed configuration. In this way, the Lagrangian variation of a quantity $A$, denoted $\Delta_{\mathrm{x}} A$, corresponds to the change of $A$ with respect to a frame that is Lie-dragged along the displacement. The relation between Eulerian and Lagrangian variations is then given by
\begin{equation}
    \Delta_\mathrm{x} = \delta + \mathcal{L}_\xi,
\end{equation}
where $\mathcal{L}_\xi$ is the Lie derivative operator along $\xi^\alpha_\mathrm{x}$ (see for example \cite{carroll_spacetime_2019} for the explicit expression of the Lie derivative of an arbitrary tensor).

In the proper frame that follows the fluid element of species $\mathrm{x}$, the velocity $u_\mathrm{x}^\alpha$ is aligned with the temporal axis. By definition of the Lagrangian variation, we have $\Delta_\mathrm{x}^{\phantom{}} u_\mathrm{x}^\alpha \propto u_\mathrm{x}^\alpha$. The proportionality factor is obtained by using the normalization of the velocity, i.e. $u_\mathrm{x}^\alpha u^\mathrm{x}_\alpha=-1$. One thus finds 
\begin{equation}
    \Delta_\mathrm{x}^{\phantom{}} u_\mathrm{x}^\alpha = \frac{1}{2}u_\mathrm{x}^\alpha u_\mathrm{x}^\mu u_\mathrm{x}^\nu \Delta_\mathrm{x}^{\phantom{}} g_{\mu\nu}.
    \label{Delta u}
\end{equation}
This formula can also be proved in a more formal way by introducing the so-called matter space~\cite{carter_covariant_1989, andersson_relativistic_2021}. The velocities $u_\mathrm{x}^\alpha$ and the current densities $n_\mathrm{x}^\alpha$ are related by
\begin{equation}
    n_\mathrm{x}^\alpha = n_\mathrm{x}^{\phantom{}} u_\mathrm{x}^\alpha,
    \label{n}
\end{equation}
where $n_\mathrm{x} = n_\mathrm{xx}$ are the number densities of each species. The Eulerian perturbations of $u^\alpha_\mathrm{x}$ and $n^\alpha_\mathrm{x}$ read
\begin{align}
    \delta u^\alpha_\mathrm{x} &= \frac{1}{2}u^\alpha_\mathrm{x}u^\mu_\mathrm{x}u^\nu_\mathrm{x} \delta g_{\mu\nu} + \big(\delta^\alpha_\nu + u^\alpha_\mathrm{x}u_\nu^\mathrm{x}\big) \big(u^\mu_\mathrm{x}\nabla_\mu \xi^\nu_\mathrm{x} - \xi^\mu_\mathrm{x}\nabla_\mu u^\nu_\mathrm{x} \big),
    \label{delta u} \\
    \delta n_\mathrm{x}^\alpha &= n_\mathrm{x}^\mu \nabla_\mu \xi_\mathrm{x}^\alpha - \xi_\mathrm{x}^\mu \nabla_\mu n_\mathrm{x}^\alpha - n_\mathrm{x}^\alpha \left(\nabla_\mu \xi_\mathrm{x}^\mu - \frac{1}{2}g_{\mu\nu}\delta g^{\mu\nu}\right). \label{delta n}
\end{align}
We clearly see that the variation of the current densities is not independent of the variation of the metric. Since we will need it later, we also give the Eulerian variation of the canonical momenta,
\begin{equation}
    \delta \pi_\alpha^\mathrm{x} = \sum_\mathrm{y} (\mathcal{B}^{\mathrm{xy}})_\alpha^{\phantom{n}\mu} \delta n_\mu^\mathrm{y} + \sum_\mathrm{y} n_\alpha^\mathrm{y} (\mathcal{H}^{\mathrm{xy}})^{\mu\nu}\delta g_{\mu\nu},
    \label{delta pi}
\end{equation}
where we have defined the following tensorial matrix elements:
\begin{align}
    (\mathcal{B}^{\mathrm{xy}})_{\mu\nu} &= \Upsilon_\mathrm{xy} \mathcal{A}_\mathrm{xy} g_{\mu\nu} - \sum_\mathrm{z,w} \Upsilon_\mathrm{xz} \Upsilon_\mathrm{yw} n_\mu^\mathrm{z} n_\nu^\mathrm{w} \frac{\partial\mathcal{A}_\mathrm{xz}}{\partial n^2_\mathrm{yw}}, \label{B} \\
    (\mathcal{H}^{\mathrm{xy}})_{\mu\nu} &= \frac{1}{2}\sum_\mathrm{z,w} \Upsilon_\mathrm{xy} \Upsilon_\mathrm{zw} n_\mu^\mathrm{z}n_\nu^\mathrm{w} \frac{\partial\mathcal{A}_\mathrm{xy}}{\partial n^2_\mathrm{zw}}.
\end{align}
It is worth noting that $(\mathcal{H}^{\mathrm{xy}})_{\mu\nu}$ is symmetric both in its tensorial indices and in its species labels, whereas $(\mathcal{B}^{\mathrm{xy}})_{\mu\nu}$ is not symmetric in general. Nevertheless, the latter is symmetric under the simultaneous exchange of tensorial and species indices.

\subsection{Constrained action principle}

We now return to the action principle and impose the constraints on the current densities.
By using \eqref{delta Lambda 2} and \eqref{delta n}, the variation of the matter action yields
\begin{equation}
    \delta(\sqrt{-g}\Lambda) = - \sqrt{-g} \left\{\frac{1}{2} \left[ \left( \Lambda - \sum_{\mathrm{x}} n^\alpha_\mathrm{x} \pi^\mathrm{x}_\alpha \right) g_{\mu\nu}+ \sum_{\mathrm{x}} \pi^\mathrm{x}_\mu n^\mathrm{x}_\nu  \right] \delta g^{\mu\nu} + \sum_{\mathrm{x}} \xi^\alpha_\mathrm{x} n^\mu_\mathrm{x} \varpi^\mathrm{x}_{\mu\alpha} \right\},
\end{equation}
where the vorticity two-form is defined as
\begin{equation}
    \varpi^\mathrm{x}_{\mu\nu} = 2 \nabla_{[\mu}^{\phantom{}} \pi^\mathrm{x}_{\nu]} = \nabla_{\mu}^{\phantom{}} \pi^\mathrm{x}_{\nu} - \nabla_{\nu}^{\phantom{}} \pi^\mathrm{x}_{\mu}.
\end{equation}
Now the independent field variables of the variational problem are assumed to be the metric
and the Lagrangian displacements. The stress-energy tensor can be inferred from the well-known formula
\begin{equation}
    T_{\mu\nu} = -\frac{2}{\sqrt{-g}} \frac{\delta (\sqrt{-g} \Lambda)}{\delta g^{\mu\nu}},
\end{equation}
and takes the form
\begin{equation}
    T^\mu_{\phantom{w} \nu} = \Psi \delta^\mu_\nu + \sum_{\mathrm{x}} n^\mu_\mathrm{x} \pi^\mathrm{x}_\nu \qquad , \qquad \Psi = \Lambda - \sum_{\mathrm{x}} n^\alpha_\mathrm{x} \pi^\mathrm{x}_\alpha,
\end{equation}
where $\Psi$ is the generalized pressure. The total variation of the action is then
\begin{equation}
    \delta \mathcal{S} = \frac{1}{16\pi} \int \mathrm{d}^4 x \sqrt{-g} \left[\left( \mathcal{R}_{\mu\nu} - \frac{1}{2} g_{\mu\nu} \mathcal{R} - 8\pi T_{\mu\nu} \right) \delta g^{\mu\nu} - 16\pi \sum_{\mathrm{x}} \xi^\alpha_\mathrm{x} n^\mu_\mathrm{x} \varpi^\mathrm{x}_{\mu\alpha}\right],
\end{equation}
where $\mathcal{R}_{\mu\nu}$ is the Ricci tensor. Requiring $\delta \mathcal{S} = 0$ yields the equations of motion,
\begin{align}
     G_{\mu \nu} \equiv \mathcal{R}_{\mu\nu} - \frac{1}{2} g_{\mu\nu} \mathcal{R} &= 8\pi T_{\mu\nu}, \label{EFE}\\
    n^\mu_\mathrm{x} \nabla_{[\mu}^{\phantom{}} \pi^\mathrm{x}_{\nu]} & = 0. \label{euler}
\end{align}
The first line corresponds to the Einstein field equations for the gravitational field, while the second represents the relativistic Euler equations governing each fluid component. These equations must be supplemented by the conservation laws $\nabla_\mu n^\mu_{\mathrm{x}} = 0$. It is worth noting that when all fluid equations in \eqref{euler} are satisfied and the fluxes are conserved, the stress-energy tensor is automatically conserved, i.e.\ $\nabla_\mu T^\mu_{\phantom{w}\nu} = 0$. Depending on the context, it may therefore be more convenient to work either with the relativistic Euler equations or with the conservation of the stress--energy tensor.

\section{The equilibrium configuration}
\label{equilibrium}

In the following, we apply Carter's formalism to model a general-relativistic multifluid compact star that is slightly deformed by the external adiabatic tidal field generated by a companion. In this section, we describe the equilibrium configuration of a isolated star. In the next section, we will perturb this configuration by introducing a small, stationary tidal field.

\subsection{The hydrostatic equations}

We consider an isolated stellar object that is static and spherically symmetric. For the treatment of rotational effects (in the one-fluid case), see~\cite{pani_tidal_2015, pani_tidal_2015-1, landry_tidal_2015, poisson_gravitomagnetic_2020}. The background metric can then be written in the usual Schwarzschild coordinates:
\begin{equation}
    \d s^2 = g_{\mu\nu}\d x^\mu \d x^\nu = -e^{\nu(r)} \d t^2 + e^{\lambda(r)} \d r^2 + r^2 \d \theta^2 + r^2 \sin^2\theta \d \phi^2.
    \label{metric_0}
\end{equation}
It is convenient to recast the $g_{rr}$ component by introducing, in analogy with the Schwarzschild solution, the mass function $m(r)$ through
\begin{equation}
    e^{-\lambda} = 1 - \frac{2m}{r}.
\end{equation}
The quantity $m(r)$ represents the mass enclosed within a spherical radius $r$.

According to Carter’s general multifluid formalism introduced earlier, each fluid constituent is associated with its own velocity. However, in a static background, all spatial components vanish, implying the absence of spatial motion. As a result, the normalization of the velocity leads to
\begin{equation}
    u^\alpha = \left(e^{-\nu/2}, 0, 0, 0 \right),
    \label{u}
\end{equation}
where the species index has been omitted since all fluids move together. The flux of each constituent can thus be written as
\begin{equation}
    n_\mathrm{x}^\alpha = n_\mathrm{x}^{\phantom{}} u^\alpha.
    \label{n background}
\end{equation}
One can readily verify that these current densities identically fulfill the conservation laws $\nabla_\mu n^\mu_{\mathrm{x}} = 0$. The canonical momenta then simplify to
\begin{equation}
    \pi_\alpha^\mathrm{x} = \pi_\mathrm{x}^{\phantom{}} u_\alpha^{\phantom{}},
    \label{pi covariant}
\end{equation}
where we have defined
\begin{equation}
    \pi_\mathrm{x} = \sum_\mathrm{y} \mathcal{A}_\mathrm{xy} \Upsilon_\mathrm{xy} n_\mathrm{y}.
    \label{pi}
\end{equation}
Thus, in a static configuration, the momenta are aligned with the conserved currents. The stress–energy tensor takes the form $T^t_{\phantom{n} t} = \Lambda$ and $T^i_{\phantom{n} j} = \Psi \delta^i_j$, where the generalized pressure is given by
\begin{equation}
    \Psi = \Lambda + \sum_\mathrm{x} n_\mathrm{x} \pi_\mathrm{x}.
    \label{Psi}
\end{equation}

The hydrostatic equations follow from the Einstein field equations together with the relativistic Euler equations, namely~\eqref{EFE} and~\eqref{euler}. The $[t,t]$ and $[r,r]$ components of Einstein’s equations yield
\begin{align}
    m' &= -4 \pi r^2 \Lambda, \\
    \nu' &= 2 \frac{m + 4 \pi r^3 \Psi}{r(r-2m)},
\end{align}
where primes denote derivatives with respect to the coordinate $r$. From the fluid equations of motion, only the radial component gives a nontrivial contribution, leading to
\begin{equation}
    \pi'_\mathrm{x} + \frac{\nu'}{2}\pi_\mathrm{x} = 0.
    \label{TOV momenta}
\end{equation}
Although these relations are naturally formulated in terms of the momenta and can be useful, it is more convenient to recast them as differential equations for the densities, since these are the fundamental variables in the multifluid description. To this end, we take the derivative of Eq.~\eqref{pi}, and after some algebra we find
\begin{equation}
    \sum_\mathrm{y} \mathcal{B}^{\mathrm{xy}} \hspace{0.05cm} n'_\mathrm{y} + \frac{\nu'}{2} \pi_\mathrm{x} = 0,
\end{equation}
where $\mathcal{B}^{\mathrm{xy}}$ can be deduced from~\eqref{B} as
\begin{equation}
    \mathcal{B}^{\mathrm{xy}} \equiv (\mathcal{B}^{\mathrm{xy}})_t^{\phantom{n}t} = \Upsilon_\mathrm{xy} \mathcal{A}_\mathrm{xy} + \sum_\mathrm{z,w} \Upsilon_\mathrm{xz} \Upsilon_\mathrm{yw} n_\mathrm{z} n_\mathrm{w} \frac{\partial\mathcal{A}_\mathrm{xz}}{\partial n^2_\mathrm{yw}}.
    \label{B tt}
\end{equation}
Note that $\mathcal{B}^{\mathrm{xy}}$ is symmetric under exchange of species indices, whereas the full tensor $(\mathcal{B}^{\mathrm{xy}})_{\mu\nu}$ is generally not symmetric. From the conservation of the stress-energy tensor, we can also infer the following useful relation:
\begin{equation}
    \Psi' = -\frac{1}{2}\nu' (\Psi - \Lambda).
    \label{psi'}
\end{equation}
This relation, however, is not independent, as it is directly linked to the relativistic Euler equations. One may similarly obtain another equation by considering the $[\theta,\theta]$ or $[\phi,\phi]$ components of the Einstein equations, but this does not provide any new information. Indeed, by virtue of the Bianchi identities, these components are entirely determined by the remaining components of the Einstein equations.

Collecting the above results, the hydrostatic equations for a multifluid compact star can be summarized as
\begin{subequations}
    \begin{align}
        m' &= -4 \pi r^2 \Lambda, \\
        \nu' &= 2 \frac{m + 4 \pi r^3 \Psi}{r(r-2m)}, \label{nu eq} \\
        0 &= \sum_\mathrm{y} \mathcal{B}^{\mathrm{xy}} n'_\mathrm{y} + \frac{\nu'}{2} \pi_\mathrm{x}. \label{euler background}
    \end{align}
    \label{hydro eqs}
\end{subequations}
These expressions constitute the generalization of the Tolman-Oppenheimer-Volkoff (TOV) equations \cite{tolman_static_1939, oppenheimer_massive_1939} to multifluid stellar models. In the particular case of two fluids, they reproduce results previously obtained in studies of the quasi-normal modes of superfluid neutron stars \cite{comer_quasinormal_1999}. For a given master function $\Lambda(n^2_\mathrm{xy})$, the system contains $N+2$ unknown functions: $\nu(r)$, $m(r)$ (or equivalently $\lambda(r)$), and the density profiles $n_\mathrm{x}(r)$. Since the number of independent differential equations is also $N+2$, the system is closed and can be integrated numerically. The number of variables may be reduced by eliminating $\nu$ through substitution of Eq.~\eqref{nu eq} into Eq.~\eqref{euler background}. To solve the system, the microphysical model determining the master function must be specified. The coupled first-order differential equations are integrated outward from the stellar center, where $m(0)=0$ and $n^\mathrm{x}(0)=n_c^\mathrm{x}$. Integration proceeds up to the radius $R$ defined by $\Psi(R)=0$, which marks the stellar surface. The total gravitational mass of the compact star is finally obtained as $M = m(R)$.

\subsection{Multifluid static thermodynamics}

In addition to the hydrostatic equations presented above, the macroscopic quantities of the system such as the energy density, pressure, and particle densities must also satisfy thermodynamical relations. Considering the static limit is particularly relevant because, even in the presence of perturbations (as we will see in the following sections), all thermodynamical quantities are evaluated on the background configuration. In this regime, all fluids are comoving, so the state functions depend solely on the number densities of each species $n_{\mathrm{x}}$, and no longer on any coupled current terms. Consequently, in the static limit one has $\Lambda = -\mathcal{E}$ and $\Psi = P$, where $\mathcal{E}$ denotes the static energy density and $P$ the pressure. In the hydrostatic equations~\eqref{hydro eqs}, one can therefore replace $(\Lambda, \Psi)$ by $(-\mathcal{E}, P)$. Note, however, that the partial derivatives must be taken using the full (non‑static) $\Lambda$, and only then evaluated on the background. In what follows, we present the key thermodynamical relations for a static multifluid system composed of $N$ species. For a more detailed discussion of thermodynamics in the multifluid hydrodynamical context, see for instance~\cite{andersson_relativistic_2021}.

Let us consider a small fluid element in the interior of a compact star. Thermodynamical quantities are defined with respect to a comoving frame attached to the fluid element. The thermodynamic state of such an element is fully characterised by a set of extensive and intensive variables. The extensive variables are the total energy $E$ (including rest-mass energy), entropy $S$, proper volume $V$, and the particle numbers $N_{\mathrm{x}}$ of each species. The intensive variables are the pressure $P$, temperature $T$, and the chemical potentials $\mu_{\mathrm{x}}$ of each species. A small change in the extensive variables induces a variation of the total energy given by
\begin{equation}
    \mathrm{d}E = T\mathrm{d}S - P\mathrm{d}V + \sum_{\mathrm{x}} \mu_{\mathrm{x}}\mathrm{d}N_{\mathrm{x}},
    \label{dE}
\end{equation}
which is the combined differential form of the first and second laws of thermodynamics. Since $E$, $S$, $V$, and $N_{\mathrm{x}}$ are extensive variables, Euler's theorem for homogeneous functions implies
\begin{equation}
    E = TS - PV + \sum_{\mathrm{x}} \mu_{\mathrm{x}} N_{\mathrm{x}}.
    \label{E}
\end{equation} 
In this formulation, the total energy $E$ acts as the fundamental thermodynamic potential, meaning that the equation of state takes the form $E = E(S, V, N_{\mathrm{x}})$, from which all thermodynamical quantities can be derived.

In hydrodynamics, it is convenient to work solely with densities, i.e. with intensive variables. For any extensive system, the number of thermodynamic variables can be reduced by one such that only intensive variables remain. We therefore introduce the following intensive quantities: the static energy density $\mathcal{E} = E/V$, the entropy density $s = S/V$ and the particle number densities $n_{\mathrm{x}} = N_{\mathrm{x}}/V$. In the multifluid context, it is convenient to treat entropy as an independent fluid component. Accordingly, the entropy density can be included among the $n_{\mathrm{x}}$, and the corresponding ``chemical'' potential is the temperature $T$. With these prescriptions, Eqs.~\eqref{dE} and \eqref{E} become 
\begin{align}
    \mathrm{d}\mathcal{E} &= \sum_{\mathrm{x}} \mu_{\mathrm{x}}\mathrm{d}n_{\mathrm{x}}, 
    \label{depsilon} \\
    \mathcal{E} &= -P + \sum_{\mathrm{x}} \mu_{\mathrm{x}} n_{\mathrm{x}}, 
    \label{epsilon}
\end{align}
from which one may also derive the Gibbs--Duhem relation,
\begin{equation}
    \mathrm{d}P = \sum_{\mathrm{x}} n_{\mathrm{x}}\mathrm{d}\mu_{\mathrm{x}}.
    \label{GD}
\end{equation}
In this formulation, the energy density $\mathcal{E} = \mathcal{E}(n_{\mathrm{x}})$ is regarded as the thermodynamic potential, and the chemical potentials follow from
\begin{equation}
    \mu_{\mathrm{x}} = \frac{\partial \mathcal{E}}{\partial n_{\mathrm{x}}},
    \label{mu}
\end{equation}
where it is understood that all other densities are held fixed.

\section{Perturbed configuration in the zero-frequency subspace}
\label{perturbation}

In a binary system, a multifluid compact star will deviate from the static and spherically symmetric configuration described in the previous section due to the tidal field generated by the companion. We assume that the bodies are sufficiently far apart that the effect can be treated as a small perturbation. The spacetime metric inside and outside the tidally perturbed object can then be written as
\begin{equation}
    \bar{g}_{\mu\nu} = g_{\mu\nu} + h_{\mu\nu},
\end{equation}
where $g_{\mu\nu}$ is the background metric and $h_{\mu\nu} = \delta g_{\mu\nu}$ is the small perturbation. We further consider the adiabatic limit, meaning that the internal (hydrodynamical) timescales are much shorter than the typical timescale over which the tidal field varies (i.e., the orbital timescale). Roughly, these characteristic timescales are $\tau_{\text{int}} \sim \sqrt{R^3/M}$ and $\tau_{\text{tide}} \sim \sqrt{d^3/M_{\text{tot}}}$, where $d$ is the orbital separation and $M_{\text{tot}}$ the total mass of the binary system. For comparable-mass systems, the internal dynamics is faster provided the ratio $R/d$ remains small. In this regime, the tidal field may effectively be regarded as stationary.

Because the background is spherically symmetric, it is convenient to expand all linearized quantities in spherical harmonics $Y_\ell^m(\theta,\phi)$. It is well known that the metric perturbation separates into two classes according to its behaviour under parity transformations, i.e., under the mapping $(\theta,\phi) \rightarrow (\pi - \theta,\, \phi + \pi)$. Thus we decompose
\begin{equation}
    h_{\mu\nu} = h_{\mu\nu}^{\text{e}} + h_{\mu\nu}^{\text{o}},
\end{equation}
where the first term corresponds to the even (polar) sector and the second to the odd (axial) sector.

As in the background configuration, the equations of motion are still given by \eqref{EFE} and \eqref{euler}, now taken at linear order. The perturbed Einstein equations are
\begin{equation}
    \delta G^\mu_{\phantom{a}\nu} = 8\pi \delta T^\mu_{\phantom{w}\nu},
    \label{delta EFE}
\end{equation}
while the perturbed Euler equations read
\begin{equation}
    \delta n_\mathrm{x}^\mu \nabla_{[\mu}^{\phantom{i}} \pi_{\nu]}^\mathrm{x} + n_\mathrm{x}^\mu \nabla_{[\mu}^{\phantom{i}} \delta\pi_{\nu]}^\mathrm{x} = 0.
    \label{delta euler}
\end{equation}
Note that it is computationally easier to work with the mixed form of the Einstein equations rather than with the purely covariant or contravariant forms. Because the background metric \eqref{metric_0} is invariant under parity transformations, it follows that all perturbed quantities—such as $\delta G^\mu_{\phantom{a}\nu}$, $\delta T^\mu_{\phantom{w}\nu}$, $\delta n_\mathrm{x}^\alpha$, and $\delta\pi_{\alpha}^\mathrm{x}$—split into polar and axial parts. As a result, the equations of motion~\eqref{delta EFE} and~\eqref{delta euler} both decouple, each leading to two independent sets of equations. We begin with the polar sector and then turn to the axial one.

\subsection{Polar sector}

To take the stationary limit in the general case, one introduces a harmonic time dependence, performs the algebra, and finally takes the zero-frequency limit. Nevertheless, it has been shown in the case of a single fluid that this procedure is equivalent, in the polar sector, to directly assuming the absence of time dependence—that is, setting all time derivatives to zero. For instance, compare the results of \cite{hinderer_tidal_2008} and \cite{damour_relativistic_2009}. The more general multifluid model considered here does not modify this conclusion. We therefore adopt this approach in the present work.

\subsubsection{The linearized metric and matter variables}

To solve the linearized equations of motion \eqref{delta EFE} and \eqref{delta euler}, we first need to specify all perturbed quantities. To fix the spurious degrees of freedom in the metric, we choose the Regge--Wheeler gauge
\cite{regge_stability_1957, thorne_non-radial_1967}:
\begin{equation}
    h_{\mu\nu}^e = \sum_{\ell,m}
    \begin{pmatrix}
        e^\nu H_0^\ell & H_1^\ell & 0 & 0 \\
        H_1^\ell & e^\lambda H_2^\ell & 0 & 0 \\
        0 & 0 & r^2 K_\ell & 0 \\
        0 & 0 & 0 & r^2 \sin^2\theta K_\ell
    \end{pmatrix}
    Y_\ell^m,
    \label{h e}
\end{equation}
where $H_0^\ell$, $H_1^\ell$, $H_2^\ell$ and $K_\ell$ are unknown functions of $r$. Because of the spherical symmetry of the background, different multipoles do not couple. Thus, the perturbations depend only on the multipole index $\ell$, and one may set $m=0$ without loss of generality. All perturbed scalar quantities are expanded in spherical harmonics as well:
\begin{align}
    \delta \Lambda &= \sum_{\ell,m} \delta \Lambda_\ell(r) Y_\ell^m,\\
    \delta \Psi    &= \sum_{\ell,m} \delta \Psi_\ell(r) Y_\ell^m,\\
    \delta n_\mathrm{x} &= \sum_{\ell,m} \delta n^\ell_\mathrm{x}(r) Y_\ell^m,
\end{align}
while the Lagrangian displacement vectors are written as
\begin{equation}
    \xi^\alpha_\mathrm{x} = \sum_{\ell,m} \left(0,W^\ell_\mathrm{x},\frac{V^\ell_\mathrm{x}}{r^2}\partial_\theta,\frac{V^\ell_\mathrm{x}}{r^2\sin^2\theta}\partial_\phi \right) Y_\ell^m,
    \label{xi e}
\end{equation}
where $W^\ell_\mathrm{x}$ and $V^\ell_\mathrm{x}$ are radial functions. Note that the components $\xi^t_\mathrm{x}$ can be set to zero due to our gauge choice. For notational simplicity, we omit the even-parity subscript, but all expressions refer to the even sector.

From these decompositions, we can infer the matter variables $\delta u^\alpha_\mathrm{x}$ and subsequently $\delta \pi_\alpha^\mathrm{x}$. Using \eqref{delta u}, one finds
\begin{equation}
    \delta u^\alpha = \frac{1}{2} \sum_{\ell,m} \left(e^{-\nu/2} H_0^\ell, 0, 0, 0\right) Y_\ell^m.
    \label{delta u e}
\end{equation}
Equation \eqref{delta u e} shows that the fluids remain strictly static in the even sector, and therefore all species share the same perturbed velocity; for this reason we omit the species index here. Using \eqref{delta pi}, one then obtains
\begin{subequations}
\begin{align}
    \delta \pi_t^\mathrm{x} &= -e^{\nu/2}\sum_{\ell,m} \left[\sum_\mathrm{y}\mathcal{B}^\mathrm{xy}\delta n^\ell_\mathrm{y}
       - \frac{1}{2}\pi_\mathrm{x} H_0^\ell\right] Y_\ell^m, \\
    \delta \pi_i^\mathrm{x} &= e^{-\nu/2}\pi_\mathrm{x}\delta^r_i \sum_{\ell,m} H_1^\ell Y_\ell^m.
\end{align}
\label{delta pi e}
\end{subequations}
The perturbed stress-energy tensor takes the form
\begin{equation}
    \delta T^\mu_{\phantom{w}\nu} = \delta \Psi\, \delta^\mu_\nu + \sum_\mathrm{x} \left(\pi_\nu^\mathrm{x} \delta n_\mathrm{x}^\mu + n_\mathrm{x}^\mu \delta \pi_\nu^\mathrm{x}\right),
    \label{T e}
\end{equation}
and, using \eqref{delta u e} and \eqref{delta pi e} together with the background relations, one finds that it reduces to
\begin{equation}
    \delta T^{\mu}_{\phantom{w}\nu} =
    \begin{pmatrix}
        \delta \Lambda & e^{-\nu}(\Psi - \Lambda) h_{01} & 0 & 0 \\
        0 & \delta \Psi & 0 & 0 \\
        0 & 0 & \delta \Psi & 0 \\
        0 & 0 & 0 & \delta \Psi
    \end{pmatrix}.
\end{equation}
These expressions fully determine the linearized metric and matter variables in the even-parity sector.

\subsubsection{The linearized equations of motion}

We now have all the necessary ingredients to derive the equations of motion for the perturbed configuration. In the single-fluid, barotropic case, it is sufficient to work solely with the Einstein equations. However, when dealing with a non-barotropic equation of state or a multifluid system, one must also use the matter equations of motion. In this context, we therefore need to consider \eqref{delta EFE} in combination with \eqref{delta euler}. This is the approach adopted in 
\cite{char_relativistic_2018, datta_effect_2020} to treat tidal deformations of superfluid neutron stars within a two-fluid model. Alternatively, one may use \eqref{delta EFE} together with the perturbed form of the stress-energy tensor conservation laws,
\begin{equation}
    \delta \left(\nabla_\mu T^\mu_{\phantom{w}\nu}\right) = 0.
\end{equation}
Both approaches are equivalent, but the latter involves only the fluid equations of motion in the background configuration. In addition, it provides a way to check the results of \cite{char_relativistic_2018, datta_effect_2020} in the special case $N=2$. It can be easily seen that the perturbed conservation laws for the currents, i.e. $\delta(\nabla_\mu n^\mu_{\mathrm{x}})=0$, are automatically satisfied.

We now turn to the linearized Einstein equations. We essentially follow the approach of Hinderer~\cite{hinderer_tidal_2008}, although the procedure 
differs
%diverges 
at certain steps due to the multifluid nature of the compact star considered. The perturbed Einstein tensor $\delta G^\mu_{\phantom{n}\nu}$ depends only on \eqref{h e} and is well known (see for example \cite{andersson_gravitational-wave_2019}). We begin with the equation $\delta G^r_{\phantom{n}t}=0$, which yields
\begin{equation}
    H_1^\ell(r) = 0.
\end{equation}
This condition implies that the perturbed metric is diagonal and, consequently, that $\delta u_r = \delta \pi_r^\mathrm{x} = 0$. Next, using $\delta G^\theta_{\phantom{n}\theta} - \delta G^\phi_{\phantom{n}\phi} = 0$, we find
\begin{equation}
    H_0^\ell(r) = H_2^\ell(r) \equiv H_\ell(r).
\end{equation}
At this stage, only two unknown functions remain in the metric perturbation: $H_\ell$ and $K_\ell$. A relation between these two functions follows from $\delta G^r_{\phantom{n}\theta} = 0$, namely
\begin{equation}
    K'_\ell = H'_\ell + \nu' H_\ell.
    \label{K'}
\end{equation}
Once $H_\ell(r)$ is known, this equation can be integrated to obtain $K_\ell(r)$. The perturbed metric is therefore fully determined by the single function $H_\ell$. It remains to derive the equations governing the fluid variables as well as a 
closed equation for $H_\ell$. As a first step, we express $\delta \Psi_\ell$ entirely in terms of $H_\ell$ by using $\delta G^\theta_{\phantom{n}\theta} + \delta G^\phi_{\phantom{n}\phi} = 16\pi \delta \Psi$, or equivalently $\delta (\nabla_\mu T^\mu_{\phantom{w}A}) = 0$ with $A \in \{\theta, \phi\}$. This leads to
\begin{equation}
    \delta \Psi_\ell = \frac{1}{2}\left(\Psi - \Lambda\right) H_\ell.
    \label{delta psi}
\end{equation}
Substituting these results in $\delta G^r_{\phantom{n}r} = 8\pi \delta \Psi$, we obtain an explicit expression for $K_\ell$ in terms of $H_\ell$ and $H'_\ell$:
\begin{equation}
    K_\ell(r) = \frac{r^2 \nu' H'_\ell + H_\ell \left[r^2 (\nu')^2 - 2 + \ell(\ell+1)e^\lambda - 8\pi r^2 e^\lambda (\Psi -\Lambda)\right]} {e^\lambda (\ell^2 + \ell - 2)}.
\end{equation}

In most of the literature on tidal effects, a cold barotropic equation of state is assumed. In this case, the pressure depends only on the energy density, i.e. $\Psi = \Psi(\Lambda)$. This greatly simplifies the determination of $\delta\Lambda_\ell$, since $\delta\Psi_\ell$ is already known from \eqref{delta psi}. Indeed, for a barotropic equation of state one has
\begin{equation}
    \delta\Psi_\ell = \frac{\d \Psi}{\d \Lambda} \delta\Lambda_\ell,
    \label{barotropic}
\end{equation}
which allows $\delta\Lambda_\ell$ to be obtained directly in terms of $H_\ell$. In the more general situation of a multifluid compact star, the equation of state—or the master function $\Lambda$—is no longer barotropic, and \eqref{barotropic} does not hold a priori. To determine $\delta\Lambda_\ell$ in full generality, we employ a particular combination of the perturbed conservation laws which, to the best of our knowledge, has not been considered previously in the context of adiabatic tides, namely %. Specifically, we consider the combination
\begin{equation}
    \partial_r\delta(\nabla_\mu T^\mu_{\phantom{a}A}) - \partial_A\delta(\nabla_\mu T^\mu_{\phantom{n}r}) = 0.
\end{equation}
This relation leads to the remarkably compact result
\begin{equation}
    \delta\Lambda_\ell = -\frac{\Lambda'}{\nu'} H_\ell \equiv -\frac{\mathcal{J}_N}{2} H_\ell,
    \label{delta Lambda perturbed}
\end{equation}
where it is convenient to introduce the quantity $\mathcal{J}_N = 2\Lambda'/\nu'$. We can proceed further by eliminating the radial derivative of $\Lambda$, so that $\mathcal{J}_N$ depends only on the momenta and the thermodynamic partial derivatives. To this end, we first express $\Lambda'$ in terms of the fundamental variables $n^2_{\mathrm{xy}}$. After some algebra, the derivative of the master function can be written as
\begin{equation}
    \Lambda' = -\sum_\mathrm{x} \pi_\mathrm{x}^{\phantom{}} n'_\mathrm{x}.
    \label{Lambda prime}
\end{equation}
Note that we have used the fact that the unperturbed configuration is static, so that the relation $n^{2}_{\mathrm{xy}} = n_{\mathrm{x}} n_{\mathrm{y}}$ holds. We now eliminate $n'_\mathrm{x}$ using the background fluid equations of motion \eqref{euler background}. These equations can be cast in the form
\begin{equation}
    \sum_\mathrm{y} \mathcal{B}^{\mathrm{xy}} n'_\mathrm{y} = -\frac{\nu'}{2} \pi_\mathrm{x}.
\end{equation}
This represents a non-homogeneous linear system of $N$ equations for the variables $n'_\mathrm{y}$. Assuming the inverse matrix~$\mathcal{B}^{-1}$ exists, we may solve for $n'_\mathrm{x}$:
\begin{equation}
    n'_\mathrm{x} = -\frac{\nu'}{2} \sum_\mathrm{y}  (\mathcal{B}^{-1})^\mathrm{xy} \pi_\mathrm{y}.
    \label{n prime}
\end{equation}
To go beyond, we need to compute the inverse matrix element $(\mathcal{B}^{-1})^\mathrm{xy}$. In principle, inverting a matrix is straightforward (even if the algebra can be tedious) when one can reliably use matrix notation. However, in our case, where the matrix $\mathcal{B}$ may have arbitrary dimension $N\times N$, we cannot use matrix notation consistently and must instead work in index notation. The general index expression for an inverse matrix element is far less familiar than its matrix counterpart; therefore, in~\ref{matrix inverse} we derive a suitable formula. From that appendix, we infer that
\begin{equation}
     (\mathcal{B}^{-1})^\mathrm{xy} = N \frac{\sum\limits_{\mathrm{x}_i,\mathrm{y}_i} \delta^{\mathrm{x}\,\mathrm{x}_1 \dots \mathrm{x}_{N-1}}_{\mathrm{y}\,\mathrm{y}_1 \dots \mathrm{y}_{N-1}}
     \mathcal{B}^{\mathrm{y}_1}_{\phantom{m}\mathrm{x}_1} \cdots \mathcal{B}^{\mathrm{y}_{N-1}}_{\phantom{m}\mathrm{x}_{N-1}}}{\sum\limits_{\mathrm{x}_i,\mathrm{y}_i}
     \delta^{\mathrm{x}_1 \dots \mathrm{x}_{N}}_{\mathrm{y}_1 \dots \mathrm{y}_{N}} \mathcal{B}^{\mathrm{y}_1}_{\phantom{m}\mathrm{x}_1} \cdots \mathcal{B}^{\mathrm{y}_{N}}_{\phantom{m}\mathrm{x}_{N}}},
     \label{inverse main}
\end{equation}
where the summation over $\mathrm{x}_i$ denotes a sum over the indices $\mathrm{x}_1, \mathrm{x}_2, \ldots, \mathrm{x}_{N-1}$ in the numerator and over $\mathrm{x}_1, \mathrm{x}_2, \ldots, \mathrm{x}_{N}$ in the denominator (and similarly for the $\mathrm{y}_i$ indices). The generalized Kronecker delta is defined in~\eqref{general delta}. We can now assemble all the pieces together. By inserting Eqs.~\eqref{Lambda prime}, \eqref{n prime}, and \eqref{inverse main} into the definition of $\mathcal{J}_N$, we obtain the following fully general expression:
\begin{equation}
    \mathcal{J}_N = N \frac{\sum\limits_{\mathrm{x},\mathrm{y}}   \sum\limits_{\mathrm{x}_i,\mathrm{y}_i} \pi_\mathrm{
    x} \pi_\mathrm{y} \delta^{\mathrm{x} \mathrm{x}_1...\mathrm{x}_{N-1}}_{\mathrm{y} \mathrm{y}_1...\mathrm{y}_{N-1}} \mathcal{B}^{\mathrm{y}_1}_{\phantom{m}\mathrm{x}_1} ...\mathcal{B}^{\mathrm{y}_{N-1}}_{\phantom{m}\mathrm{x}_{N-1}} }{\sum\limits_{\mathrm{x}_i,\mathrm{y}_i} \delta^{\mathrm{x}_1...\mathrm{x}_{N}}_{ \mathrm{y}_1...\mathrm{y}_{N}} \mathcal{B}^{\mathrm{y}_1}_{\phantom{m}\mathrm{x}_1} ...\mathcal{B}^{\mathrm{y}_{N}}_{\phantom{m}\mathrm{x}_{N}}}.
    \label{J}
\end{equation}
Once the master function $\Lambda$ is specified, the quantity $\mathcal{J}_N$ can be computed directly from this expression, since $\pi_\mathrm{x}$ and $\mathcal{B}^{\mathrm{xy}}$ are fully determined by $\Lambda$ and its thermodynamic partial derivatives; see Eqs.~\eqref{A}, \eqref{pi}, and \eqref{B tt}.

Having obtained an explicit expression for $\mathcal{J}_N$, and hence for $\delta\Lambda_\ell$, we may now return to the Einstein equations in order to derive a single differential equation governing the remaining degree of freedom, namely $H_\ell$. This equation is obtained by considering the specific combination
$\delta G^t_{\phantom{n}t} - \delta G^r_{\phantom{n}r} = 8\pi (\delta\Lambda - \delta\Psi)$. Making use of all the relations derived previously, we finally arrive at
\begin{equation}
    H''_\ell + H'_\ell \left[4\pi re^\lambda (\Psi+\Lambda) + \frac{1+e^\lambda}{r}\right] + H_\ell \left[4\pi e^\lambda (9\Psi - 5\Lambda + \mathcal{J}_N) - (\nu')^2 - \frac{e^\lambda}{r^2} \ell(\ell+1) \right] = 0.
    \label{eq H}
\end{equation}
This equation plays the central role in the determination of the gravitoelectric-type Love numbers of a multifluid compact object. When compared to the single-fluid result \cite{hinderer_tidal_2008, damour_relativistic_2009}, one finds that the overall structure remains remarkably similar. The primary distinction arises from the appearance of the term $\mathcal{J}_N$, which, a priori, seems quite different from its one-fluid counterpart and appears to incorporate a much broader range of microphysical effects. A detailed discussion of $\mathcal{J}_N$ will be given in Sec.~\ref{entrainement}, where several questions will be clarified. An important point to keep in mind is that, since the equation was derived within linear perturbation theory, all the quantities inside the brackets in \eqref{eq H} must be evaluated in the background configuration. This means that whenever $\Lambda$ or $\Psi$ appear, they must be evaluated in the static limit, where they simply reduce to (minus) the energy density and the pressure, respectively. The term $\mathcal{J}_N$ must also be evaluated in the background configuration. Since it is essentially composed of partial derivatives of the master function, one must first compute all derivatives using the general (non-static) form of $\Lambda(n^{2}_{\mathrm{xy}})$, and only at the end take the static limit.

\subsubsection{Applications of the general formalism}

Exploiting this general framework, we will explicitly recover the well-known single-fluid equation originally derived by Hinderer \cite{hinderer_tidal_2008}, as well as reproduce the two-fluid equations of Refs.~\cite{char_relativistic_2018, datta_effect_2020}. Finally, we extend the formalism by providing an explicit expression for $\mathcal{J}_N$ in the case of three fluid species.\\

\paragraph{Single fluid problem}

In this case, all thermodynamic quantities depend only on the density $n$. The master function and the generalized pressure then reduce to (minus) the energy density and to the pressure, respectively:
\begin{equation}
    \Lambda(n) = -\mathcal{E}(n) \qquad , \qquad \Psi(n) = P(n).
\end{equation}
Here, this relation always holds, even in the non‑static configuration. The equation of state is therefore barotropic, and one may write $P=P(\mathcal{E})$. It is straightforward to verify that the general expression~\eqref{J} reduces to
\begin{equation}
    \mathcal{J}_1 = \frac{\mu^2}{\mathcal{B}},
\end{equation}
where all species indices have been dropped since there is only a single fluid. Here, we have denoted $\pi_\mathrm{x}$ by $\mu_\mathrm{x}$. Using the definitions given in~\eqref{pi} and~\eqref{B tt}, together with the thermodynamic relation \eqref{mu}, one finds that $\mu$ is indeed the chemical potential, and
\begin{equation}
    \mathcal{B} = \frac{\mathrm{d} \mu}{\mathrm{d} n}.
\end{equation}
This identification is natural, since the momenta $\pi_\mu^\mathrm{x}$ are both canonically and thermodynamically conjugate to the currents $n^\mu_\mathrm{x}$. Using the fundamental Euler relation \eqref{epsilon}, $P + \mathcal{E}=\mu n$, one obtains
\begin{equation}
    \mathcal{B} = \frac{\mu}{n}\frac{\d P}{\d \mathcal{E}}.
\end{equation}
We finally arrive at
\begin{equation}
    \mathcal{J}_1 = \frac{P+\mathcal{E}}{\d P/\d \mathcal{E}}.
    \label{J1}
\end{equation}
This is the well-known result originally obtained by Hinderer~\cite{hinderer_tidal_2008}.\\

\paragraph{Two-fluid problem}

In the case $N=2$, we label the two fluid species by `1' and `2'. Using the symmetry property of $\mathcal{B}^{1}_{\phantom{a}2}$, we obtain
\begin{equation}
    \mathcal{J}_2 = \frac{\pi_1^2 \mathcal{B}^2_{\phantom{a}2} + \pi_2^2 \mathcal{B}^1_{\phantom{a}1} - 2 \pi_1 \pi_2 \mathcal{B}^1_{\phantom{a}2}}{\mathcal{B}^1_{\phantom{a}1}\mathcal{B}^2_{\phantom{a}2} - \bigl(\mathcal{B}^1_{\phantom{a}2}\bigr)^2}.
\end{equation}
This expression coincides with the result obtained in Refs.~\cite{char_relativistic_2018, datta_effect_2020} in the context of the two-fluid model of cold superfluid neutron stars. As emphasized in the introduction, the present formalism is more general and can also be applied to other physical situations that have not yet been explored in the literature.\\

\paragraph{Three-fluid problem}

We now consider the case $N=3$ and label the fluid species by `1', `2', and `3'. Proceeding along the same lines as in the previous cases, one can show, after some algebra, that
\begin{equation}
    \begin{split}
        \mathcal{J}_3 &= \Big \{\pi^2_1 \left[\mathcal{B}^2_{\phantom{a}2}\mathcal{B}^3_{\phantom{a}3} - (\mathcal{B}^2_{\phantom{a}3})^2 \right] + \pi_2^2 \left[\mathcal{B}^1_{\phantom{a}1}\mathcal{B}^3_{\phantom{a}3} - (\mathcal{B}^1_{\phantom{a}3})^2 \right] + \pi^2_3 \left[\mathcal{B}^1_{\phantom{a}1}\mathcal{B}^2_{\phantom{a}2} - (\mathcal{B}^1_{\phantom{a}2})^2 \right]\\
        & - 2\pi_1 \pi_2 \left[\mathcal{B}^3_{\phantom{a}3}\mathcal{B}^1_{\phantom{a}2} - \mathcal{B}^2_{\phantom{a}3}\mathcal{B}^1_{\phantom{a}3} \right] - 2\pi_1 \pi_3 \left[\mathcal{B}^2_{\phantom{a}2}\mathcal{B}^1_{\phantom{a}3} - \mathcal{B}^1_{\phantom{a}2}\mathcal{B}^2_{\phantom{a}3} \right] - 2\pi_2 \pi_3 \left[\mathcal{B}^1_{\phantom{a}1}\mathcal{B}^2_{\phantom{a}3} - \mathcal{B}^1_{\phantom{a}2}\mathcal{B}^1_{\phantom{a}3} \right] \Big\}\\
        & \times \Big \{\mathcal{B}^1_{\phantom{a}1}\mathcal{B}^2_{\phantom{a}2}\mathcal{B}^3_{\phantom{a}3} - \mathcal{B}^1_{\phantom{a}1}(\mathcal{B}^2_{\phantom{a}3})^2 - \mathcal{B}^3_{\phantom{a}3}(\mathcal{B}^1_{\phantom{a}2})^2 - \mathcal{B}^2_{\phantom{a}2}(\mathcal{B}^1_{\phantom{a}3})^2 + 2\mathcal{B}^1_{\phantom{a}2}\mathcal{B}^2_{\phantom{a}3}\mathcal{B}^1_{\phantom{a}3}\Big \}^{-1},
    \end{split}
\end{equation}
where we have also used the fact that the matrix $\mathcal{B}^{\mathrm{xy}}$ is symmetric. This is the first time that this factor has been calculated for a three-fluid system.

\subsection{Axial sector}

We now consider the axial part of the perturbations. In the polar sector, assuming stationary perturbations is equivalent to treating time-dependent perturbations and then taking the zero-frequency limit at the end. However, in the axial sector this equivalence no longer holds, as reported in Ref.~\cite{pani_magnetic_2018} for the single-fluid case. The same behaviour is expected for a multifluid star. We therefore consider both approaches, which correspond to the cases of static and irrotational fluids. As shown in Refs.~\cite{gupta_relativistic_2021,gupta_effect_2023}, both assumptions are relevant for characterizing the gravitomagnetic tidal responses long before and after a mode resonance. To construct the most general framework, we begin by allowing all quantities to depend on time.

\subsubsection{The linearized metric and matter variables}

As before, we first write down all the linearized quantities and then compute the equations of motion. The odd-parity metric perturbation, in the Regge-Wheeler gauge~\cite{regge_stability_1957, thorne_non-radial_1967}, reads
\begin{equation}
    h_{\mu\nu}^o = \sum_{\ell,m}
    \begin{pmatrix}
        0 & 0 & -\frac{h_0^\ell}{\sin \theta}\partial_\phi & \sin\theta \, h_0^\ell \, \partial_\theta \\
        0 & 0 & -\frac{h_1^\ell}{\sin \theta}\partial_\phi & \sin\theta \, h_1^\ell \, \partial_\theta \\
        -\frac{h_0^\ell}{\sin \theta}\partial_\phi & -\frac{h_1^\ell}{\sin \theta}\partial_\phi & 0 & 0 \\
        \sin\theta \, h_0^\ell \, \partial_\theta & \sin\theta \, h_1^\ell \, \partial_\theta & 0 & 0
    \end{pmatrix}
    Y_\ell^m,
    \label{h 0}
\end{equation}
where $h_0^\ell$ and $h_1^\ell$ are unknown functions of $t$ and $r$. We write the Lagrangian displacements as
\begin{equation}
    \xi^\alpha_\mathrm{x} = \sum_{\ell,m} \left(0,0,-\frac{U^\ell_\mathrm{x}}{r^2\sin\theta}\partial_\phi,\frac{U^\ell_\mathrm{x}}{r^2 \sin\theta}\partial_\theta \right) Y_\ell^m,
    \label{xi o}
\end{equation}
where $U_\mathrm{x}^\ell=U_\mathrm{x}^\ell(t,r)$. It is important to note that under parity transformations, scalar quantities have even parity. Therefore, in the axial sector, all scalar field perturbations vanish identically.

Following the same procedure as in the even-parity case, one computes the perturbed velocities, momenta, and stress-energy tensor from \eqref{h 0} and \eqref{xi o}. Using \eqref{delta u}, one then obtains
\begin{equation}
    \delta u_\mathrm{x}^\alpha = \sum_{\ell,m} \left(0,0,-\frac{e^{-\nu/2}}{r^2\sin\theta} \dot{U}^\ell_\mathrm{x}\partial_\phi,\frac{e^{-\nu/2}}{r^2 \sin\theta}\dot{U}^\ell_\mathrm{x}\partial_\theta \right) Y_\ell^m,
    \label{delta u o}
\end{equation}
where a dot denotes a partial derivative with respect to time. From \eqref{pi}, the perturbed momenta are then given by
\begin{equation}
    \delta \pi^\mathrm{x}_\alpha = \sum_{\ell,m} \sum_\mathrm{y} \Upsilon_\mathrm{xy}\mathcal{A}_\mathrm{xy}n_\mathrm{y} \left(0,0,-\frac{e^{-\nu/2}}{\sin\theta} \left(h_0^\ell+\dot{U}^\ell_\mathrm{x}\right)\partial_\phi,e^{-\nu/2} \sin\theta \left(h_0^\ell+\dot{U}^\ell_\mathrm{x}\right)\partial_\theta \right) Y_\ell^m.
    \label{delta pi o}
\end{equation}
In the axial sector, the stress-energy tensor \eqref{T e} simplifies to
\begin{equation}
    \delta T^\mu_{\phantom{w}\nu} = \sum_\mathrm{x} \left(\pi_\nu^\mathrm{x} \delta n_\mathrm{x}^\mu + n_\mathrm{x}^\mu \delta \pi_\nu^\mathrm{x} \right),
    \label{T o}
\end{equation}
and one can verify that only four components of the perturbed stress-energy tensor are nonvanishing: $\delta T^t_{\phantom{n}\theta}$, $\delta T^t_{\phantom{n}\phi}$, $\delta T^\theta_{\phantom{n}t}$, and $\delta T^\phi_{\phantom{n}t}$. The first two are given by
\begin{subequations}
    \begin{align}
        \delta T^t_{\phantom{n}\theta} &= -\frac{e^{-\nu}}{\sin\theta}\sum_\mathrm{x}n_\mathrm{x}\pi_\mathrm{x} \sum_{\ell,m}\left(h_0^\ell+\dot{U}_\mathrm{x}^\ell \right) \partial_\phi Y_\ell^m, \\
        \delta T^t_{\phantom{n}\phi} &= e^{-\nu}\sin\theta\sum_\mathrm{x}n_\mathrm{x}\pi_\mathrm{x} \sum_{\ell,m}\left(h_0^\ell+\dot{U}_\mathrm{x}^\ell \right) \partial_\theta Y_\ell^m.
\end{align}
\end{subequations}
The remaining components can be obtained by contraction with the metric.

\subsubsection{The linearized equations of motion}

We can now write down the perturbed Einstein equations. In the odd-parity sector, there are only three independent equations. They can be obtained by considering the components $\delta G^\theta_{\phantom{n}\theta}=0$, $\delta G^r_{\phantom{n}\theta}=0$, and $\delta G^t_{\phantom{n}\theta}=8\pi \delta T^t_{\phantom{n}\theta}$. This leads to
\begin{subequations}
    \begin{align}
        e^{-\nu}\dot{h}_0 - e^{-\lambda}h_1 - \frac{h_1}{r^2}\left[4\pi r^3 (\Psi+\Lambda)+2m \right] &= 0, \label{eq1}\\[0.8ex]
        e^{-\nu}\left(\dot{h}'_0-\ddot{h}_1 \right) - \frac{2}{r}e^{-\nu}\dot{h}_0 - \frac{h_1}{r^2}(\ell-1)(\ell+2) &= 0, \label{eq2} \\[0.8ex]
        e^{-\lambda}\left(h''_0 - \dot{h}'_1\right) - 4\pi r (\Psi-\Lambda) \left(h'_0-\dot{h}_1 \right) - \frac{2}{r}e^{-\lambda}\dot{h}_1 \nonumber \\
        - \frac{h_0}{r^3} \left[8\pi(\Psi-\Lambda)r^3 + \ell(\ell+1)r - 4m\right] - 16\pi \sum_\mathrm{x} n_\mathrm{x}\pi_\mathrm{x}\dot{U}_\mathrm{x}&= 0. \label{eq3}
\end{align}
\label{eqs axial}
\end{subequations}
The multipole index $\ell$ has been omitted to lighten the notation. In addition to these equations, we also invoke the perturbed Euler equations. Inspecting \eqref{delta euler}, it is straightforward to see that the $\nu=t$ and $\nu=r$ components are trivially satisfied. The remaining two components imply the following evolution equations for the fluid momenta:
\begin{equation}
    \partial_t \delta \pi_\mu^\mathrm{x} = 0.
    \label{dot delta pi}
\end{equation}
We therefore have a total of $N+3$ equations—namely \eqref{eqs axial} together with \eqref{dot delta pi}—for $N+2$ unknown functions: $h_0^\ell$, $h_1^\ell$, and $U_\mathrm{x}^\ell$. In fact, only $N+2$ equations are independent due to the Bianchi identities, but the key point is that the system provides the necessary number of equations to determine all variables. We now consider the two regimes discussed at the beginning of this subsection.\\

\paragraph{Static fluids}

We first follow essentially the same procedure as in the even-parity sector. We assume from the outset that the perturbations are stationary, i.e. all time derivatives vanish. Under this assumption, we have
\begin{equation}
    \dot{h}_0^\ell = \dot{h}_1^\ell = \dot{U}^\ell_\mathrm{x} = 0.
\end{equation}
The conditions $\dot{U}^\ell_{\mathrm{x}} = 0$ imply that the fluids remain purely static; see~\eqref{delta u o}. This leads to substantial simplifications. Equation \eqref{eq2} immediately yields $h_1^\ell = 0$, which in turn makes \eqref{eq1} automatically satisfied. The relativistic Euler equations \eqref{dot delta pi} are also trivially satisfied. At this point, the only remaining degree of freedom in the axial sector is~$h_\ell \equiv h_0^\ell$. Inserting the above simplifications into \eqref{eq3} gives
\begin{equation}
    h_\ell'' - 4\pi r e^\lambda (\Psi-\Lambda)\, h'_\ell + e^\lambda h_\ell \left[\frac{4m}{r^3} - 8\vartheta\pi(\Psi-\Lambda) - \frac{\ell(\ell+1)}{r^2}\right] = 0,
    \label{master eq static}
\end{equation}
where we have introduced the parameter $\vartheta = 1$. The usefulness of this definition will become clear later. This is the equation that must be solved to determine the gravitomagnetic tidal Love numbers for purely \emph{static} fluids. It is identical to the equation derived in \cite{binnington_relativistic_2009, pani_magnetic_2018}, with the replacement of the energy density and pressure by $-\Lambda$ and $\Psi$, respectively. However, because we consider linear perturbations, these latter quantities must be evaluated on the static background. As a result, the master equation for $h_\ell$ in the static case takes exactly the same form as in the barotropic, single-fluid limit.\\

\paragraph{Irrotational fluids}

We now return to the general system \eqref{eqs axial} and \eqref{dot delta pi}, without imposing staticity, and follow the second procedure outlined earlier. We assume that all perturbed quantities are time dependent with a harmonic profile. That is, for any function $f(t,r)$, we take $f(t,r) = f(r)e^{i\Omega_m t}$, where $\Omega_m$ is the mode frequency. We perform the calculation with this ansatz and subsequently take the limit $\Omega_m \to 0$. From \eqref{dot delta pi}, we immediately obtain
\begin{equation}
    \delta\pi_\mu^\mathrm{x} = 0.
    \label{delta pi 0}
\end{equation}
Using \eqref{delta pi o}, these conditions imply
\begin{equation}
    \sum_\mathrm{x} \Upsilon_\mathrm{xy}\mathcal{A}_\mathrm{xy} n_\mathrm{x} \left(h^\ell_0 + \dot{U}_\mathrm{x}^\ell\right) = 0.
    \label{eq}
\end{equation}
A more convenient form is obtained by multiplying this equation by $n_\mathrm{y}$ and summing over $\mathrm{y}$. Making use of \eqref{Psi}, we arrive at
\begin{equation}
    \sum_\mathrm{x} \pi_\mathrm{x}^{\phantom{}} n_\mathrm{x}^{\phantom{}} \dot{U}_\mathrm{x}^\ell = -(\Psi - \Lambda) h_0^\ell.
\end{equation}
This relation allows us to rewrite the last term in \eqref{eq3}. We then substitute the harmonic time dependence and finally take the zero-frequency limit. After simplification, we obtain
\begin{equation}
    h_\ell'' - 4\pi re^\lambda (\Psi-\Lambda) h'_\ell + e^\lambda h_\ell \left[\frac{4m}{r^3} - 8\vartheta\pi (\Psi-\Lambda) - \frac{\ell(\ell+1)}{r^2}\right] = 0,
    \label{master eq irrot}
\end{equation}
where $h_\ell \equiv h_0^\ell$ and, in this case, $\vartheta = -1$. This is the equation governing the calculation of the gravitomagnetic Love numbers for \emph{irrotational} fluids. Comparing \eqref{master eq irrot} with the static-fluid result \eqref{master eq static} shows that the two procedures are inequivalent: the equations differ by the sign of the parameter $\vartheta$. As in the static case, the master function and the generalized pressure are evaluated on the background and therefore reduce to $-\mathcal{E}$ and $P$, respectively. Consequently, Eq.~\eqref{master eq irrot} takes the same form as the equation obtained in the single-fluid case in~\cite{landry_gravitomagnetic_2015, pani_magnetic_2018}. Our result differs from the equations presented in 
\cite{damour_relativistic_2009, datta_effect_2020}, because those works employ a different independent variable, namely
\begin{equation}
    \psi_\ell = \frac{h_1^\ell}{r} e^{(\nu - \lambda)/2}.
\end{equation}
Using \eqref{eq1}, one can express $\dot{h}_\ell$ in terms of $\psi_\ell$. Substituting this relation into \eqref{eq2} and taking the zero-frequency limit yields an equation that is different in form but equivalent, now written in terms of $\psi_\ell$.

We have asserted that \eqref{master eq irrot} is the perturbation equation governing fluids in an irrotational state. At this stage, the statement is not yet obvious, so let us examine the nature of the fluids. Within the general multifluid formalism, a given species is said to be irrotational if its vorticity 2-form vanishes, i.e.
\begin{equation}
     \varpi_{\mu\nu}^\mathrm{x} = 2 \nabla_{[\mu}^{\phantom{}} \pi^\mathrm{x}_{\nu ]}=0.
     \label{irrot}
\end{equation}
In the linearized regime, this becomes
\begin{equation}
    \delta\varpi_{\mu\nu}^\mathrm{x} = 2 \nabla_{[\mu}^{\phantom{}} \delta\pi^\mathrm{x}_{\nu ]}=0.
    \label{irrot2}
\end{equation}
This condition appears to differ from those used in \cite{pani_magnetic_2018, landry_gravitomagnetic_2015}. Let us clarify the connection. In \cite{pani_magnetic_2018}, a single-fluid system is said to be irrotational if the vorticity vector defined as
\begin{equation}
    \omega^\alpha = \frac{1}{2}\varepsilon^{\alpha\mu\rho\sigma}u_\mu \nabla_\rho u_\sigma,
    \label{omega 1}
\end{equation}
vanishes, where $\varepsilon^{\alpha\mu\rho\sigma}$ is the completely antisymmetric Levi-Civita tensor. For comparison with the multifluid framework, the vorticity can be recast as
\begin{equation}
    \omega^\alpha_\mathrm{x} = \frac{1}{4\pi^2_\mathrm{x}} \varepsilon^{\mu\alpha\rho\sigma} \pi_\mu^\mathrm{x} \varpi^\mathrm{x}_{\rho\sigma},
    \label{omega 2}
\end{equation}
with $\pi^2_\mathrm{x} = -\pi_\mu^\mathrm{x}\pi_\mathrm{x}^\mu$. In the single-fluid limit, \eqref{omega 2} is equivalent to \eqref{omega 1}. Hence, if \eqref{irrot} holds (the fluids are irrotational), then $\omega^\alpha_\mathrm{x}=0$ follows automatically; the converse, however, does not necessarily hold. Therefore, the condition $\omega^\alpha_\mathrm{x}=0$ is not, by itself, sufficient to guarantee full irrotationality of the multifluid system. As an illustration, in the background one finds $\omega^\alpha=0$ while $\varpi_{tr}\neq 0$, which is consistent with a static configuration. The previous discussion concerned the background relation, whereas here we are interested in the perturbed quantities. Perturbing \eqref{omega 2} shows that, in general, no straightforward relation exists between $\delta \omega^\alpha_\mathrm{x}$ and $\delta\varpi^\mathrm{x}_{\mu\nu}$. In particular, the irrotationality condition \eqref{irrot2} is not equivalent to imposing $\delta \omega^\alpha_\mathrm{x}=0$. Therefore, the vanishing of the vorticity vector cannot be used as the defining criterion for irrotationality.

Still in the single-fluid context, \cite{landry_gravitomagnetic_2015} adopts the criterion that an irrotational fluid satisfies
\begin{equation}
    \delta u_\alpha = 0.
\end{equation}
In a multifluid setting, a natural generalization is
\begin{equation}
    \delta \pi_\mu^\mathrm{x}=0.
    \label{irrot LP}
\end{equation}
Imposing \eqref{irrot LP} indeed implies $\delta \varpi^\mathrm{x}_{\mu\nu} = 0$, so $\delta \pi_\mu^\mathrm{x}=0$ is a sufficient condition for irrotationality. However, it is not necessary: one may have $\delta \pi_\mu^\mathrm{x}\neq 0$ while still satisfying $\delta \varpi^\mathrm{x}_{\mu\nu}=0$. In particular, $\delta \pi_\mu^\mathrm{x} = \nabla_\mu \delta\Phi_\mathrm{x}$, where $\Phi_\mathrm{x}$ denotes a scalar field (as required locally for a superfluid, in which case $\Phi_\mathrm{x}$ is proportional to the phase of the condensate) satisfies the irrotationality condition (this condition holds locally for a superfluid but not necessarily  
globally if quantized vortices are present). Thus, the condition discussed in \cite{landry_gravitomagnetic_2015} is not the most general criterion for irrotational flow.

With these clarifications in place, we now show that the fluids are irrotational in our setup. From the equations of motion \eqref{delta pi 0}, we have $\delta\pi_\mu^\mathrm{x} = 0$, which immediately enforces the vanishing of the vorticity 2-form for each species via~\eqref{irrot2}. This completes the proof that our configuration is irrotational, thereby justifying the use of \eqref{master eq irrot} in this case.

\section{Tidal Love numbers and deformabilities}
\label{Love numbers}

In the preceding sections, we established the equilibrium structure of a static and spherically symmetric multifluid compact star, before introducing perturbations and analysing how its fluid constituents respond. However, up to this point, we have not specified the nature of these perturbations. In the present context, they arise from the stationary tidal field generated by the companion. This tidal field can be decomposed into even-parity (electric) and odd-parity (magnetic) components. We denote the electric and magnetic tidal moments of order $\ell$ by $\mathscr{E}_L$ and $\mathcal{M}_L$, respectively, where the multi-index $L = i_1 i_2 \dots i_\ell$ follows the notation of \cite{damour_general-relativistic_1991}. These tidal moments perturb the multifluid body by exciting the various mass multipole moments $\mathcal{Q}_L$ and current multipole moments $S_L$. At linear order, and in the adiabatic limit, the tidal fields and the induced multipole moments are related by
\begin{align}
    \mathscr{E}_L &= \lambda_\ell \mathcal{Q}_L, \label{lambda l} \\
    \mathcal{M}_L &= \sigma_\ell S_L, \label{sigma l}
\end{align}
where the linear coefficients are the electric (magnetic) tidal deformabilities $\lambda_\ell$ ($\sigma_\ell$) of order $\ell$. Since these are dimensional quantities, it is more convenient to introduce their dimensionless counterparts. For this purpose, it is useful to define the electric and magnetic tidal Love numbers as
\begin{align}
k_\ell &= \frac{1}{2}(2\ell-1)!!\frac{\lambda_\ell}{R^{2\ell+1}}, \\
j_\ell &= 4(2\ell-1)!!\frac{\sigma_\ell}{R^{2\ell+1}} .
\end{align}
Note that our normalization of the gravitomagnetic Love numbers differs from that of Ref.~\cite{damour_relativistic_2009} and follows that of Ref.~\cite{perot_role_2021}. The dimensionless tidal deformabilities, which directly enter the gravitational-wave phase, are given by
\begin{align}
\Lambda_\ell &= \frac{2}{(2\ell-1)!!} k_\ell \left(\frac{R}{M}\right)^{2\ell+1}, \label{Lambda l} \\
\Sigma_\ell &= \frac{1}{4(2\ell-1)!!} j_\ell \left(\frac{R}{M}\right)^{2\ell+1}. \label{Sigma l}
\end{align}
These potentially observable coefficients do not depend on the chosen normalization of the Love numbers.

\subsection{Gravitoelectric tidal Love numbers}

The electric Love numbers are calculated as follows; see \cite{hinderer_tidal_2008, damour_relativistic_2009, binnington_relativistic_2009}. The first step is to solve the perturbation equation \eqref{eq H} outside the star. In this region, we have $\Psi=\Lambda=0$. The perturbation equation can be solved analytically, yielding the general solution
\begin{equation}
    H_\ell(x) = a_{\ell}^P P_{\ell 2}(x) + a_\ell^Q Q_{\ell 2}(x),
    \label{H out}
\end{equation}
where $x=r/M - 1$, and $P_{\ell 2}(x)$ and $Q_{\ell 2}(x)$ are the normalized associated Legendre functions of the first and second kinds, respectively. Far from the star, as $x \rightarrow \infty$, these functions behave as $P_{\ell 2}(x)\sim x^\ell$ and $Q_{\ell 2}(x)\sim x^{-(\ell+1)}$. To determine the constants $a_\ell^P$ and $a_\ell^Q$, we match the interior and exterior solutions at the stellar surface. A convenient way to perform this matching is to introduce the function
\begin{equation}
    y_\ell(r)=\frac{rH'_\ell}{H_\ell},
    \label{y e}
\end{equation}
and impose the continuity condition $y_\ell^\text{in}(R)=y_\ell^\text{out}(R)$ (see \cite{aranguren_review_2024} for a detailed discussion of the matching conditions). This allows the constants to be expressed directly in terms of $y_\ell \equiv y_\ell^\text{in}(R)$. The final step in obtaining an explicit expression for $k_\ell$ is to compare \eqref{H out} with the multipole expansion of the metric outside the perturbed compact star. In the body’s local asymptotic rest frame (i.e., asymptotically mass-centered Cartesian coordinates), the $[t,t]$ component of the metric at large $r$ takes the form
\begin{equation}
    \frac{1 + h_{tt}}{2} = \frac{(2\ell - 1)!!}{\ell!} n^L r^{-(\ell+1)}\mathcal{Q}_L + \frac{1}{\ell!} n^L r^\ell \mathscr{E}_L,
    \label{H multipole}
\end{equation}
where $n^L=x^L/r^\ell$ is the unit radial multi-vector. We see that the metric contains a decreasing part, corresponding to the gravitational field generated by the star, and an increasing part, corresponding to the external tidal field. By comparing the asymptotic form of \eqref{H out} with \eqref{H multipole}, and using \eqref{lambda l}, one extracts the electric Love numbers. Explicit expressions up to $\ell=5$ can be found in \cite{perot_role_2021}. Since future third-generation gravitational-wave detectors are expected to measure mainly the quadrupolar and octupolar contributions, we quote only the expressions for $k_2$ and $k_3$:
\begin{align}
    k_2 &= \frac{8}{5} C^5 (1-2C)^2 \big[ 2(y_2-1)C - y_2 + 2 \big] \times \Big\{ 2C \big[ 4(y_2+1)C^4 + 2(3y_2-2)C^3 - 2(11y_2-13)C^2 \nonumber \\
    &+ 3(5y_2-8)C - 3(y_2-2) \big] + 3(1-2C)^2 \big[ 2(y_2-1)C-y_2+2 \big] \log(1-2C) \Big\}^{-1}, \label{k2} \\[1.5ex]
    k_3 &= \frac{8}{7} C^7 (1-2C)^2 \big[ 2(y_3-1)C^2 - 3(y_3-2)C + y_3 - 3 \big] \times \Big\{ 2C \big[ 4(y_3+1)C^5 + 2(9y_3-2)C^4 \nonumber \\
    & - 20(7y_3-9)C^3 + 5(37y_3-72)C^2 - 45(2y_3-5)C + 15(y_3-3) \big] \nonumber \\
    &+ 15(1-2C)^2 \big[ 2(y_3-1)C^2 - 3(y_3-2)C + y_3 - 3 \big] \log(1-2C) \Big\}^{-1}, \label{k3}
\end{align}
where $C = M/R$ is the compactness parameter. By solving the hydrostatic equations \eqref{hydro eqs} together with the perturbation equation~\eqref{eq H} inside the star, and then evaluating~\eqref{y e} at the surface, we can compute the electric Love numbers using the previous formulas. The explicit expressions \eqref{k2} and \eqref{k3} are the same as in the single-fluid case; the only difference arises from the numerical values of $y_\ell$.

\subsection{Gravitomagnetic tidal Love numbers}

The calculation of magnetic-type Love numbers follows the same strategy; see \cite{damour_relativistic_2009, binnington_relativistic_2009}. The exterior solution of the perturbation equations \eqref{master eq static} and \eqref{master eq irrot} is
\begin{equation}
    h_\ell(\tilde{x}) = b_\ell^Q F_\ell^Q(\tilde{x}) + b_\ell^P F_\ell^P(\tilde{x}),
    \label{h out}
\end{equation}
where $\tilde{x}=r/M$, and $F_\ell^Q(\tilde{x})$ and $F_\ell^P(\tilde{x})$ are hypergeometric functions. Far from the star, these functions behave as $F_\ell^Q(\tilde{x})\sim \tilde{x}^{\ell+1}$ and $F_\ell^P(\tilde{x})\sim \tilde{x}^{-\ell}$. The constants in \eqref{h out} are determined by matching the quantity
\begin{equation}
    \mathcal{Y}_\ell(r)=\frac{r h'_\ell}{h_\ell},
    \label{y o}
\end{equation}
across the interior and exterior solutions at the surface. In this way, the constants can be expressed in terms of $\mathcal{Y}_\ell \equiv \mathcal{Y}_\ell^\text{in}(R)$. As in the electric case, the connection to the Love numbers is established by comparison with the metric multipole expansion:
\begin{equation}
    \sum_{\ell,m} \ell(\ell+1)\frac{h_\ell}{r^2}Y_\ell^m = \frac{4(2\ell-1)!!}{(\ell-1)!} n^L r^{-(\ell-2)}S_L + \frac{1}{(\ell-1)!} n^L r^{\ell-1}\mathcal{M}_L.
\end{equation}
Using the asymptotic form of \eqref{h out} together with Eq.~\eqref{sigma l} yields the magnetic Love numbers. The quadrupolar and octupolar ones are
\begin{align}
    j_2 &= \frac{24}{5} C^5 \big[ 2(\mathcal{Y}_2-2)C - \mathcal{Y}_2 + 3 \big] \times \Big\{ 2C \big[ 2(\mathcal{Y}_2+1)C^3 + 2\mathcal{Y}_2 C^2 + 3(\mathcal{Y}_2-1)C - 3(\mathcal{Y}_2-3) \big] \nonumber \\
    &+ 3 \big[ 2(\mathcal{Y}_2-2)C - \mathcal{Y}_2 + 3 \big] \log(1-2C) \Big\}^{-1}, \label{j2} \\[1.5ex]
    j_3 &= \frac{64}{21} C^7 \big[ 8(\mathcal{Y}_3-2)C^2 - 10(\mathcal{Y}_3-3)C + 3(\mathcal{Y}_3-4) \big] \times \Big\{ 2C \big[ 4(\mathcal{Y}_3+1)C^4 + 10\mathcal{Y}_3C^3 + 30(\mathcal{Y}_3-1)C^2 \nonumber \\
    &- 15(7\mathcal{Y}_3-18)C + 45(\mathcal{Y}_3-4) \big] + 15 \big[ 8(\mathcal{Y}_3-2)C^2 - 10(\mathcal{Y}_3-3)C + 3(\mathcal{Y}_3-4) \big] \log(1-2C) \Big\}^{-1}. \label{j3}
\end{align}
Magnetic Love numbers up to $\ell=5$ are given in \cite{perot_role_2021}. These numbers are numerically computed using the latter formulas, where $\mathcal{Y}_\ell$ is obtained by first solving the perturbation equation~\eqref{master eq static} or~\eqref{master eq irrot} inside the compact star, and then evaluating~\eqref{y o} at the surface. As in the electric sector, the expressions for the magnetic Love numbers are the same as in the single-fluid case. The multifluid effects enter only through the numerical values of $\mathcal{Y}_\ell$.

\section{Imprint of entrainment effect on the tidal responses}
\label{entrainement}

In the previous sections, we have developed the general formalism needed to compute the tidal deformabilities of a multifluid compact star. The natural question that arises is how this multifluid behaviour affects the tidal deformations, and in particular how mutual entrainment influences them. As mentioned in the introduction, there are discrepancies in the existing literature regarding the case of cold superfluid neutron stars, as treated within a two-fluid model. Here, we clarify these issues in the more general context of an $N$‑component system. We then apply this general analysis to the cases of superfluid neutron stars and dark matter admixed compact objects.

\subsection{The analytical multifluid equation of state}

To investigate how entrainment affects the perturbation equations \eqref{eq H}, \eqref{master eq static} and \eqref{master eq irrot}, and thus the resulting tidal deformations, we must go one step further and introduce an explicit form of the master function. We adopt the following analytical expression:
\begin{equation}
    \Lambda(n^2_\mathrm{xy})=\frac{1}{2}\sum_{k=0}^\infty \sum_\mathrm{z,w} \mathcal{X}_k^\mathrm{zw} \big(n^2_\mathrm{zw}-n_\mathrm{z}n_\mathrm{w} \big)^k,
    \label{analytical master}
\end{equation}
where the coefficients $\mathcal{X}_k^\mathrm{xy}$ depend only on the number density of each species, i.e.\ $\mathcal{X}_k^\mathrm{xy} = \mathcal{X}_k^\mathrm{xy}(n_\mathrm{z})$, and are symmetric in the fluid indices. This expression generalises the form introduced in \cite{andersson_oscillations_2002} for the two-fluid case. Let us outline the motivation behind this structure. Equation~\eqref{analytical master} is rather general, as it represents a power-series expansion of $\Lambda$ around the point $n^2_\mathrm{xy}=n_\mathrm{x}n_\mathrm{y}$. Under the assumption that $\Lambda$ is analytic (in the mathematical sense), the above expression is exact. In addition, this form is particularly useful in regimes where the three‑velocities are small. Indeed, in a sufficiently small region of spacetime, one can locally write (see \cite{andersson_oscillations_2002})
\begin{equation}
    n^2_\mathrm{xy} = n_\mathrm{x}n_\mathrm{y}\left[\frac{1-\delta_{ij}u^i_\mathrm{x} u^j_\mathrm{y}}{\sqrt{(1-v_\mathrm{x}^2)(1-v^2_\mathrm{y})}} \right],
\end{equation}
where $v_\mathrm{x}$ is defined by $v_\mathrm{x}=\delta_{ij}u^i_\mathrm{x}u_\mathrm{x}^j$. Consequently, in the limit of small three‑velocities, i.e.\ $|u^i_\mathrm{x}|\ll1$, one simply recovers $n^2_\mathrm{xy} \approx n_\mathrm{x}n_\mathrm{y}$. The analytical form of $\Lambda$ can therefore be interpreted as an expansion in small spatial velocities. This viewpoint is particularly convenient because all thermodynamic quantities must be evaluated in the static background, where $u^i_\mathrm{x}=0$. As a result, even though the full expansion \eqref{analytical master} is considered, many terms vanish when evaluated on the background. With this in mind, we may rewrite the master function in a more transparent way:
\begin{equation}
    \Lambda(n^2_\mathrm{xy}) = \mathcal{X}_0 + \frac{1}{2}\sum_{k=1}^\infty \sum_\mathrm{z,w} \mathcal{X}_k^\mathrm{zw} \big(n^2_\mathrm{zw}-n_\mathrm{z}n_\mathrm{w} \big)^k,
    \label{analytical master2}
\end{equation}
where we have defined
\begin{equation}
    \mathcal{X}_0 = \frac{1}{2}\sum_{\mathrm{z,w}}\mathcal{X}_0^{\mathrm{zw}}.
\end{equation}
Evaluated on the equilibrium configuration, this simply gives $\Lambda = \mathcal{X}_0$, so that $\mathcal{X}_0$ corresponds to minus the static energy density $\mathcal{E}$. The non‑static contributions, and hence the entrainment effect, are encoded in the coefficients $\mathcal{X}_k^\mathrm{xy}$ for $k>0$. Since self‑entrainment is excluded, we impose that $\mathcal{X}_k^\mathrm{xx}=0$.

From the analytical expression of $\Lambda$, we can now calculate all the thermodynamic quantities. This requires evaluating partial derivatives of $n_\mathrm{xy}$ with respect to $n_\mathrm{zw}$. If the matrix element $n_\mathrm{xy}$ were arbitrary, with no symmetry property, we would have the usual formula
\begin{equation}
    \frac{\partial n_\mathrm{xy}}{\partial n_\mathrm{zw}} = \delta^\mathrm{x}_\mathrm{z}\, \delta^\mathrm{y}_\mathrm{w}.
\end{equation}
However, $n_\mathrm{xy}$ is symmetric, so this relation is no longer valid. In particular, it would yield $\partial n_\mathrm{xy}/\partial n_\mathrm{yx} = 0$, whereas the correct value should clearly be one. We therefore require a derivative rule that returns one when differentiating an object with respect to itself and zero otherwise, is manifestly symmetric under the exchange of both index pairs, and remains valid in all cases. This leads to
\begin{equation}
    \frac{\partial n_\mathrm{xy}}{\partial n_\mathrm{zw}} = \delta^\mathrm{(x}_\mathrm{z}\, \delta^\mathrm{y)}_\mathrm{w}\, \big(2 - \delta^\mathrm{x}_\mathrm{y}\big).
\end{equation}
The first factor accounts for the symmetric nature of $n_\mathrm{xy}$, while the second ensures the correct behaviour when the two indices coincide. One may easily verify that this is the only choice satisfying the criteria listed above.

With this subtlety noted, we can now evaluate the thermodynamic quantities. Using \eqref{A} and~\eqref{B tt}, we find after some algebra that
\begin{align}
    \mathcal{A}_\mathrm{xy} &= \mathcal{C}_\mathrm{xy} - \delta_\mathrm{xy} \mathcal{T_\mathrm{x}}, \label{A analytic}\\
    \mathcal{B}_\mathrm{xy} &= -\frac{\partial^2 \mathcal{X}_0}{\partial n_\mathrm{x}\partial n_\mathrm{y}} - \frac{1}{2} \sum_{k=1}^\infty \sum_{\mathrm{z,w}} \frac{\partial^2\mathcal{X}_k^\mathrm{zw}}{\partial n_\mathrm{x} \partial n_\mathrm{y}}(n^2_\mathrm{zw}-n_\mathrm{z}n_\mathrm{w})^k, \label{B analytic}
\end{align}
where we have defined
\begin{align}
    \mathcal{C}_\mathrm{xy} &= -\sum_{k=1}^\infty k \mathcal{X}_k^\mathrm{xy} (n^2_\mathrm{xy}-n_\mathrm{x}n_\mathrm{y})^{k-1}, \\
    \mathcal{T}_\mathrm{x} &= \frac{1}{4n_\mathrm{x}}\sum_{k=0}^\infty \sum_{\mathrm{z,w}} \frac{\partial\mathcal{X}_k^\mathrm{zw}}{\partial n_\mathrm{x}}(n^2_\mathrm{zw}-n_\mathrm{z}n_\mathrm{w})^k + \frac{1}{2} \sum_{\mathrm{z}} \frac{n_\mathrm{z}}{n_\mathrm{x}} \mathcal{C}_\mathrm{xz}.
\end{align}
We see from \eqref{A analytic} that a natural separation between diagonal and off-diagonal terms emerges. The matrix $\mathcal{C}_\mathrm{xy}$ has vanishing diagonal elements and fully captures the entrainment effect. Even in the static limit, $\mathcal{C}_\mathrm{xy} \neq 0$, which shows that entrainment contributes to the matrix $\mathcal{A}_\mathrm{xy}$. However, we see from \eqref{B analytic} that substantial simplifications occur and only the first term survives when evaluated in the background. Therefore, we obtain the important result that, in the background,
\begin{equation}
    \mathcal{B}_\mathrm{xy} = \frac{\partial \mu_\mathrm{x}}{\partial n_\mathrm{y}} = \frac{\partial \mu_\mathrm{y}}{\partial n_\mathrm{x}}.
    \label{B BG}
\end{equation}
We have used the fact that $\mathcal{X}_0 = -\mathcal{E}$ together with the thermodynamical relation \eqref{mu}. We may also verify that $\pi_\mathrm{x}$ reduce to the chemical potentials in the static limit. Indeed, using \eqref{pi}, we find that
\begin{equation}
    \pi_\mathrm{x} = -\frac{\partial \mathcal{X}_0}{\partial n_\mathrm{x}} - \frac{1}{2} \sum_{k=1}^\infty \sum_{\mathrm{z,w}} \frac{\partial\mathcal{X}_k^\mathrm{zw}}{\partial n_\mathrm{x}}(n^2_\mathrm{zw}-n_\mathrm{z}n_\mathrm{w})^k.
\end{equation}
It is then straightforward to see that $\pi_\mathrm{x}=\mu_\mathrm{x}$ in the background.

\subsection{Tidal deformabilities do not depend on the entrainment effect}

As already discussed in the previous sections, the thermodynamical quantities entering the hydrostatic equations~\eqref{hydro eqs} and the perturbation equations \eqref{eq H}, \eqref{master eq static} and \eqref{master eq irrot} are evaluated in the background. This implies that $\Lambda$ and $\Psi$ reduce to $-\mathcal{E}$ and $P$, respectively. All the quantities appearing in these differential equations then have a clear interpretation, except for $\mathcal{J}_N$, which enters the gravitoelectric perturbation equation. A priori, it is not straightforward to compute this quantity in the static limit by simply inserting \eqref{B BG} directly into \eqref{J}. We therefore proceed in a slightly different way.

We return to the hydrostatic equations and note that \eqref{TOV momenta} can be recast as
\begin{equation}
    \mu_\mathrm{x} e^{\nu/2} = \mu_\mathrm{x}^c = \mathrm{constant},
    \label{mu const}
\end{equation}
where $\mu_\mathrm{x}^c = \mu_\mathrm{x}(r=0)$. Thus, we have $N$ relations describing the variation of the chemical potentials throughout the interior of the compact star. Since all species satisfy this relation, it is possible to choose a reference chemical potential among the $N$ components, say $\mu_1$, and express all others as functions of it. Dividing \eqref{mu const} by $\mu_{1} e^{\nu/2} = \mu_{1}^c$, we obtain
\begin{equation}
    \mu_\mathrm{x} = \alpha_\mathrm{x} \mu_1,
    \label{mu equations}
\end{equation}
where we have defined $\alpha_\mathrm{x} = \mu_\mathrm{x}^c / \mu_1^c$, a set of constants. Since the chemical potentials depend on the densities, $\mu_\mathrm{x} = \mu_\mathrm{x}(n_\mathrm{y})$, relations \eqref{mu equations} provide a system of $N-1$ independent equations for the $N$ densities (the equation for $\mathrm{x}=1$ is trivially satisfied). The densities can therefore be expressed as functions of a single reference density, say $n_1$: $n_\mathrm{x} = n_\mathrm{x}(n_1)$. We will use these dependencies with respect to reference quantities to recast $\mathcal{J}_N$ in a more convenient way. We can already rewrite \eqref{J} as
\begin{equation}
    \mathcal{J}_N = \mu_1^2 \sum_\mathrm{x,y}  (\mathcal{B}^{-1})^\mathrm{xy}\alpha_\mathrm{x}\alpha_\mathrm{y}.
    \label{JN}
\end{equation}
Inspecting \eqref{B BG}, we see that $\mathcal{B}^\mathrm{xy}$ evaluated in the background is the Jacobian matrix of $\mu_\mathrm{x}(n_\mathrm{y})$. We may therefore construct its inverse,
\begin{equation}
    (\mathcal{B}^{-1})^\mathrm{xy} = \frac{\partial n_\mathrm{x}}{\partial \mu_\mathrm{y}},
    \label{B-1}
\end{equation}
where $n_\mathrm{x}$ is viewed as a functional of $\mu_\mathrm{y}$, i.e.\ $n_\mathrm{x}=n_\mathrm{x}(\mu_\mathrm{y})$. Using \eqref{mu equations}, each density $n_\mathrm{x}$ can be expressed as a function of the reference chemical potential $\mu_1$ alone. Hence,
\begin{equation}
    \frac{\mathrm{d} n_\mathrm{x}}{\mathrm{d} \mu_1} = \sum_\mathrm{y} \frac{\partial n_\mathrm{x}}{\partial \mu_\mathrm{y}} \frac{\mathrm{d} \mu_\mathrm{y}}{\mathrm{d} \mu_1} = \sum_\mathrm{y} (\mathcal{B}^{-1})^\mathrm{xy} \alpha_\mathrm{y},
    \label{dn dmu}
\end{equation}
where we used \eqref{mu equations}, which implies $\mathrm{d}\mu_\mathrm{x} = \alpha_\mathrm{x} \mathrm{d}\mu_1$, and \eqref{B-1}. Since each $n_\mathrm{x}$ can also be viewed as a function of $n_1$ only, we have
\begin{equation}
    \frac{\mathrm{d} n_\mathrm{x}}{\mathrm{d} \mu_1}
    = \frac{\mathrm{d} n_\mathrm{x}}{\mathrm{d} n_1}
      \frac{\mathrm{d} n_1}{\mathrm{d} \mu_1}.
\end{equation}
We now substitute this relation into \eqref{dn dmu} and multiply the result by $\mu_1^2 \alpha_\mathrm{x}$. After summing over $\mathrm{x}$, comparison with~\eqref{JN} gives
\begin{equation}
    \mathcal{J}_N
    = \frac{\mu_1^2}{\mathrm{d}\mu_1/\mathrm{d}n_1}
      \sum_\mathrm{x} \alpha_\mathrm{x} \frac{\mathrm{d} n_\mathrm{x}}{\mathrm{d} n_1}.
    \label{JN2}
\end{equation}
To go further, we recast the thermodynamical relations \eqref{depsilon}, \eqref{epsilon} and \eqref{GD} introduced earlier:
\begin{align}
    \mathrm{d} \mathcal{E} & = \mu_1 \mathrm{d} n_1 \sum_\mathrm{x} \alpha_\mathrm{x} \frac{\mathrm{d} n_\mathrm{x}}{\mathrm{d} n_1},
       \label{dE bis}\\
    \mathrm{d} P & = \mathrm{d}\mu_1 \sum_\mathrm{x} \alpha_\mathrm{x} n_\mathrm{x},
       \label{dP bis} \\
    \mathcal{E} & = -P + \mu_1 \sum_\mathrm{x} \alpha_\mathrm{x} n_\mathrm{x}.
    \label{E+P}
\end{align}
Using \eqref{dE bis}, we rewrite \eqref{JN2} as
\begin{equation}
    \mathcal{J}_N \frac{\mathrm{d}\mu_1}{\mathrm{d} n_1}
    = \mu_1 \frac{\mathrm{d} \mathcal{E}}{\mathrm{d} n_1}.
\end{equation}
Multiplying by $n_\mathrm{x}\alpha_\mathrm{x}$, summing over $\mathrm{x}$, and making use of \eqref{dP bis} and \eqref{E+P}, we finally arrive at
\begin{equation}
    \mathcal{J}_N
    = \frac{\mathcal{E} + P}{\mathrm{d}P/\mathrm{d}\mathcal{E}}.
    \label{J final}
\end{equation}
This is exactly the same result as in the single-fluid, barotropic case; see \eqref{J1}.

With this result, we can now rewrite all the differential equations that we need. The hydrostatic equations for the background are
\begin{align}
    m' &= 4 \pi r^2 \mathcal{E}, \\
    \nu' &= 2 \frac{m + 4 \pi r^3 P}{r(r-2m)},\\
    0 &= \sum_\mathrm{y} \frac{\partial \mu_\mathrm{x}}{\partial n_\mathrm{y}} n'_\mathrm{y} + \frac{\nu'}{2} \mu_\mathrm{x},
\end{align}
while the perturbation equations are
\begin{align}
    H''_\ell + H'_\ell \left[4\pi re^\lambda (P-\mathcal{E}) + \frac{1+e^\lambda}{r}\right]
    + H_\ell \left[4\pi e^\lambda \left(9P + 5\mathcal{E} + \frac{\mathcal{E}+P}{\mathrm{d}P/\mathrm{d}\mathcal{E}} \right)
    - (\nu')^2 - \frac{e^\lambda}{r^2}\ell(\ell+1) \right] &= 0,\\[1.5ex]
    h_\ell'' - 4\pi re^\lambda (P+\mathcal{E}) h'_\ell + e^\lambda h_\ell \left[\frac{4m}{r^3} - 8\vartheta\pi(P+\mathcal{E}) - \frac{\ell(\ell+1)}{r^2}\right] &= 0.
\end{align}
The key point is that all thermodynamic quantities entering these equations are the static energy density, the pressure, and their derivatives, all fully determined by the coefficients $\mathcal{X}_0^\mathrm{xy}$ only. Thus, no trace of the entrainment coefficients $\mathcal{X}_k^\mathrm{xy}$ with $k>0$ appears: setting all these coefficients to zero at the very beginning of the calculation would lead to the same result. Consequently, entrainment has no impact on the hydrostatic and perturbation equations. This implies that entrainment has no effect on the global structure of the star and on the tidal deformabilities, in both parity sectors and for any multipolar order. This is an important result: regardless of the potentially complicated microphysical description of the compact star, only the first term in the expansion \eqref{analytical master2} is required to describe adiabatic tides. This conclusion still holds if the star consists of a multifluid core surrounded by a single-fluid envelope.

While the tidal deformabilities do not depend on entrainment, they still depend on the densities of the different constituents because the equation of state is not, in general, barotropic. However, as shown in this section, all densities can be expressed as functions of a single reference density through the background relations~\eqref{mu const}. This induces an effective on-shell barotropicity, allowing the derivative $\mathrm{d}P/\mathrm{d}\mathcal{E}$ to be consistently defined. Still, the relation between $\mathcal{E}$ and $P$ depends on the specification of the parameters $\alpha_\mathrm{x}$ defined in~\eqref{mu equations}, and this will affect the numerical values of the tidal deformabilities. We may therefore summarise the main result of this section as follows: the adiabatic tidal responses are uniquely determined by the static equation of state $\mathcal{E}(n_\mathrm{x})$, independently of entrainment effect.

\subsection{Superfluid neutron stars}

A few ten to hundred years after their formation, neutron stars are expected to become cold enough for the existence of superfluid neutrons in their inner crust and core, and superconducting protons in their core. The presence of such superfluid phases in neutron stars is supported by observations of the cooling of young isolated neutron stars (see, e.g., Ref.~\cite{zhao_verification_2025} and references therein), transiently accreting neutron stars (see, e.g., Ref.~\cite{allard_gapless_2024}) and pulsar frequency glitches (see Ref.~\cite{antonopoulou_pulsar_2022} for a review). A fully realistic macroscopic description of superfluid neutron stars involves several independent fluid components \cite{rau_relativistic_2020}. However, a minimal model based on the two-fluid formalism is often employed~\cite{langlois_differential_1998}. For the present discussion, we assume that the core is composed of neutrons, protons, and electrons, whereas the inner crust contains nuclei, electrons, and neutrons. Since the star is assumed to be cold, there is no entropy fluid. In the minimal model, all charged species are taken to move together due to electromagnetic coupling, forming a single electrically neutral ``normal'' component. Thus, in both the core and the inner crust, we have a neutron superfluid and a normal fluid containing all other species. This latter component is often referred to as the ``proton'' fluid and includes protons and electrons in the core, and nuclei and electrons in the inner crust. A cold superfluid neutron star can therefore be described by a two-fluid model. All the results developed in the previous sections can therefore be applied to this case.

The effects of superfluidity on tidal deformations within such a model were first investigated in \cite{char_relativistic_2018, datta_effect_2020}. These works reported that solving the two-fluid hydrostatic and perturbation equations numerically leads to tidal deformabilities differing by up to $20\%$ from the one-fluid (nonsuperfluid) case. In these studies, it was assumed that the background configuration is in beta equilibrium, i.e.\ $\mu_n = \mu_p$, where $\mu_n$ and $\mu_p$ denote the chemical potentials of the neutron and ``proton'' fluids, respectively (the electron contribution being included in the latter). Note that this condition satisfies the equation of motion \eqref{mu equations} when the constant $\alpha$ is set to unity. Let us now compare these numerical findings with the conclusion of the previous subsection. In the minimal two-fluid model, the static energy density has the form $\mathcal{E}=\mathcal{E}(n_n,n_p)$, where $n_n$ and $n_p$ are the neutron and proton number densities. In fact, this energy density reduces to the familiar one-fluid, barotropic form widely used in the literature. From equation~\eqref{mu equations}, and in particular from the beta-equilibrium condition, one may express $n_p$ as a function of $n_n$, and thus rewrite the energy density as $\mathcal{E}=\mathcal{E}(n_b)$, with $n_b = n_n + n_p$ the baryon number density. According to the results of the previous subsection, tidal deformabilities depend only on the static energy density and not on the entrainment. Since superfluid effects enter exclusively through the entrainment coefficients $\mathcal{X}_k^\mathrm{xy}$ with $k>0$, it follows that superfluidity leaves no imprint on adiabatic tidal responses. All relevant physics is determined solely by the cold, beta-equilibrated, barotropic equation of state. This shows that the results of \cite{char_relativistic_2018, datta_effect_2020} most likely arise from numerical issues. Our conclusions instead agree with the reasoning presented in \cite{passamonti_dynamical_2022}, where it was anticipated that superfluidity has no impact on adiabatic tidal responses when the background is in beta equilibrium.

In fact, our conclusions are even more general: beta equilibrium is not required. In a non–beta-equilibrated configuration, the only change appears in the static energy density. The equation of state will differ, and the resulting tidal deformabilities will change accordingly, but superfluidity will still leave no imprint on the tidal responses. Similarly, one may include additional degrees of freedom, such as an entropy fluid or a superfluid hyperonic component. The energy density will change, but superfluidity itself will continue to leave no trace in the gravitational-wave signal. The general conclusion is therefore the following: the tidal deformabilities do not depend on superfluidity of some components. Only the choice of the static energy density $\mathcal{E}(n_\mathrm{x})$ matters for adiabatic tides. This will still be the case if the presence of the crust is taken into account, both as a single fluid or as an elastic domain.

\subsection{Dark matter admixed compact stars}

One way to incorporate dark matter into the theoretical description of the Universe is to allow it to be present inside compact stars, in particular in the cores of neutron stars~\cite{bramante_dark_2024}. The standard approach is to model such systems within a two-fluid formalism, wherein one component is baryonic matter and the other is dark matter~\cite{ciarcelluti_have_2011, leung_tidal_2022}. In general, the corresponding master function takes the form $\Lambda(n_b^2, n_d^2, n_{bd}^2)$, where the labels $b$ and $d$ denote baryonic and dark matter, respectively. It is common to write
\begin{equation}
    \Lambda(n_b^2, n_d^2, n_{bd}^2)
    = \Lambda_b(n_b^2) + \Lambda_d(n_d^2) + \Lambda_{\mathrm{int}}(n_b^2, n_d^2, n_{bd}^2),
\end{equation}
where the first two terms represent (minus) the energy densities of the baryonic and dark-matter fluids, while the last term encodes the interaction between them. To the best of our knowledge, all calculations of the tidal deformabilities of dark matter admixed compact stars assume a vanishing interaction term~\cite{leung_tidal_2022, barbat_comprehensive_2024, mukhopadhyay_compact_2017, das_dark_2022, zhang_gw170817_2022}. Under this assumption, the equations decouple and the analysis simplifies. However, we cannot exclude the possibility of weak interactions between dark and baryonic matter. In any case, entrainment (if present) does not affect the tidal deformabilities and therefore cannot be constrained from gravitational-wave measurements of the adiabatic tides of dark matter admixed compact stars. In practice, if one wishes to include an interaction term, it is sufficient to consider one that depends only on the number densities and not on the coupled current, i.e. $\Lambda_{\mathrm{int}} = \Lambda_{\mathrm{int}}(n_b^2, n_d^2)$. Indeed, the part of the interaction that depends on the coupled current does not contribute to the tidal deformabilities.

\section{Conclusion}

This work was motivated by the upcoming advent, potentially within the next decade, of third-generation ground-based gravitational-wave observatories. Since theoretical modeling is expected to constitute the dominant source of systematic uncertainty, increasingly accurate models incorporating richer physics will be required. With this objective in mind, the present paper has sought to contribute to this broader effort. We have developed a comprehensive general‑relativistic framework to describe the adiabatic tidal deformations of multifluid compact stars composed of an arbitrary number $N$ of interacting constituents. Building on Carter’s multifluid variational formalism, we derived the full set of hydrostatic equations as well as the linear perturbation equations governing both gravitoelectric and gravitomagnetic tidal responses. This approach extends far beyond the standard single–perfect-fluid model and provides a unified description applicable to systems such as superfluid neutron stars, dark matter admixed objects, hot compact stars, and other multifluid configurations.

The key result of this analysis is the demonstration that the nondissipative mutual entrainment effect---although an essential aspect of multifluid systems---does not affect the adiabatic tidal responses. This remains true even if the outer layer of the star consists of a single fluid or a solid. In other words, entrainment has no impact on the tidal deformabilities in both the polar and axial sectors and for any multipolar order. This result carries several noteworthy physical implications, two of which we have discussed. First, it implies that the superfluid nature of neutron stars leaves no imprint on the gravitational-wave signal from tidally deformed binary neutron stars as long as internal oscillation modes are not excited. This clarifies certain apparent contradictions between Refs.~\cite{char_relativistic_2018, datta_effect_2020} and Ref.~\cite{passamonti_dynamical_2022}. Second, it shows that any entrainment between baryons and dark matter does not influence the tidal Love numbers of dark matter admixed compact stars. Hence, ignoring such coupling---as is done in all current calculations of the adiabatic tides of such objects---does not actually introduce any approximation. These conclusions can be extended to any physically relevant multifluid compact-star model. This means that entrainment of any kind between the constituting fluids of a compact star does not leave any imprint on the adiabatic tides, and therefore cannot be probed from the gravitational waves emitted by such objects in binaries during their early inspiral.

It is important to emphasize that the absence of any imprint of the entrainment applies only to the setup considered here: a static, spherically symmetric body perturbed by an adiabatic external tidal field. In more general situations, this conclusion will not necessarily remain true. For instance, if the background is an axisymmetric rotating star or if the tidal field becomes dynamical rather than adiabatic, the entrainment effect may no longer cancel exactly. For example, superfluidity has a small but non-vanishing role in dynamical tides \cite{passamonti_dynamical_2022}. However, the assumption of a static background and an adiabatic tidal field remains very justified in the early inspiral.

Several extensions of this work are possible. One could, for instance, allow some of the fluids to carry electric charge or incorporate dissipative processes. Carter’s multifluid framework can naturally accommodate all these aspects; see Ref.~\cite{andersson_relativistic_2021} for a review. Depending on the specific compact‑star model under consideration, such effects may become relevant. It would be interesting to investigate whether adding these additional physical aspects would modify the key result of this paper, namely that entrainment has no impact on adiabatic tides. It is also possible to relax the restriction to stationary perturbations and address the full dynamical problem. Carter's formalism can then be employed to study the quasi-normal modes of multifluid compact stars. Although this has been explored in the two-fluid case for superfluid neutron stars~\cite{comer_quasinormal_1999, andersson_oscillations_2002}, it has, to our knowledge, never been generalized to an arbitrary number of fluid species.

\section*{Acknowledgements}

E.C. is a FRIA grantee of the F.R.S.-FNRS (Belgium). N.C. is a senior research associate of the F.R.S.-FNRS. This work also receives support from the FWO (Belgium) and the F.R.S.-FNRS under the Excellence of Science (EOS) programme (project No. 40007501).

\appendix

\renewcommand{\thesection}{Appendix \Alph{section}}
\renewcommand{\theequation}{\Alph{section}.\arabic{equation}}

\section{Inverse matrix in tensorial notation}
\label{matrix inverse}

Let us consider an arbitrary square matrix of dimension $n \times n$, whose components are denoted by $F^a_{\phantom{w}b}$. In this appendix, Latin letters label matrix indices and therefore run from $1$ to $n$. More generally, we may regard $F$ as a tensor, in which case the placement of indices becomes important and the Einstein summation convention applies. The goal of this appendix is to compute the elements of the inverse matrix, $(F^{-1})^a_{\phantom{w}b}$.

We begin by recalling a basic result from linear algebra, namely that the inverse matrix can be written as
\begin{equation}
    \left(F^{-1}\right)^a_{\phantom{w}b} = \frac{\left(\text{adj}(F)\right)^a_{\phantom{w}b}}{\text{det}(F)},
    \label{inverse F}
\end{equation}
where $\text{adj}(F)$ denotes the adjugate matrix, i.e., the transpose of the cofactor matrix, and $\text{det}(F)$ is the determinant. It is well known that the determinant of a matrix can be expressed in the following form~\cite{cramlet_general_1926}:
\begin{equation}
    \text{det}(F) = \frac{1}{n!} \delta_{b_1...b_n}^{a_1...a_n} F^{b_1}_{\phantom{m}a_1}...F^{b_n}_{\phantom{m}a_n},
    \label{determinant F}
\end{equation}
where the generalized Kronecker delta is defined as
\begin{equation}
    \delta_{b_1...b_n}^{a_1...a_n} = \epsilon^{a_1...a_n} \epsilon_{b_1...b_n} = \sum_{\sigma\in S_n} \text{sgn}(\sigma) \delta^{a_1}_{b_{\sigma(1)}}...\delta^{a_n}_{b_{\sigma(n)}},
    \label{general delta}
\end{equation}
with $\epsilon^{a_1...a_n}$ denoting the completely antisymmetric Levi-Civita symbol. The sum runs over all permutations $\sigma$ of the permutation group $S_n$ of order $n$. Having established an explicit expression for the determinant, it remains to derive the corresponding expression for the adjugate matrix. This can be achieved by exploiting a remarkable property of the determinant, namely 
\begin{equation}
    \left(\text{adj}(F)\right)^a_{\phantom{w}b} = \frac{\partial(\text{det}(F))}{\partial F^b_{\phantom{w}a}}.
\end{equation}
A proof of this identity in tensorial notation is given in Refs.~\cite{glendenning_compact_1996, grinfeld_introduction_2013}. Substituting the expression \eqref{determinant F} into the above relation and performing some algebra, one finds
\begin{equation}
     \left(\text{adj}(F)\right)^a_{\phantom{m}b} = \frac{1}{(n-1)!}  \delta^{a a_1...a_{n-1}}_{b b_1...b_{n-1}} F^{b_1}_{\phantom{m}a_1} ...F^{b_{n-1}}_{\phantom{m}a_{n-1}}.
     \label{adj}
\end{equation}
Combining Eqs.~\eqref{determinant F} and \eqref{adj} with~\eqref{inverse F}, we obtain the final expression for the elements of the inverse matrix:
\begin{equation}
    \left(F^{-1}\right)^a_{\phantom{w}b} = n \frac{\delta^{a a_1...a_{n-1}}_{b b_1...b_{n-1}} F^{b_1}_{\phantom{m}a_1} ...F^{b_{n-1}}_{\phantom{m}a_{n-1}}}{\delta_{b_1...b_n}^{a_1...a_n} F^{b_1}_{\phantom{m}a_1}...F^{b_n}_{\phantom{m}a_n}}.
    \label{inverse2}
\end{equation}

While the above expression is valid for arbitrary tensors, we now apply this result to the matrix 
%within the framework of the main text. While the above expression is valid for arbitrary tensors, in the specific case of the matrix 
$\mathcal{B}$ defined in \eqref{B tt}, recalling that the position of the fluid indices is irrelevant here. For this reason, the Einstein summation convention is not used and all sums are written explicitly. The inverse of this matrix is thus explicitly given by 
\begin{equation}
    \left(\mathcal{B}^{-1}\right)^\mathrm{xy} = N \frac{\sum\limits_{\mathrm{x}_i,\mathrm{y}_i} \delta^{\mathrm{x} \mathrm{x}_1...\mathrm{x}_{N-1}}_{\mathrm{y} \mathrm{y}_1...\mathrm{y}_{N-1}} \mathcal{B}^{\mathrm{y}_1}_{\phantom{m}\mathrm{x}_1} ...\mathcal{B}^{\mathrm{y}_{N-1}}_{\phantom{m}\mathrm{x}_{N-1}} }{\sum\limits_{\mathrm{x}_i,\mathrm{y}_i} \delta^{\mathrm{x}_1...\mathrm{x}_{N}}_{ \mathrm{y}_1...\mathrm{y}_{N}} \mathcal{B}^{\mathrm{y}_1}_{\phantom{m}\mathrm{x}_1} ...\mathcal{B}^{\mathrm{y}_{N}}_{\phantom{m}\mathrm{x}_{N}}}.
    \label{inverse appendix}
\end{equation}
Here, the summation over $\mathrm{x}_i$ runs over $\mathrm{x}_1$, $\mathrm{x}_2$, and so forth, up to $\mathrm{x}_{N-1}$ in the numerator and $\mathrm{x}_N$ in the denominator, with an analogous notation for the $\mathrm{y}_i$ indices.

\bibliography{references}

\end{document}